\documentclass[journal=jacsat,manuscript=article]{achemso}

\usepackage[version=3]{mhchem} 
\usepackage{multirow}
\usepackage[colorlinks=true,linkcolor=blue,citecolor=blue,urlcolor=blue]{hyperref}
\SectionNumbersOn
\usepackage{amssymb}

\newcommand*{\citen}{}
\DeclareRobustCommand*{\citen}[1]{%
  \begingroup
    \romannumeral-`\x 
    \setcitestyle{numbers}%
    \cite{#1}%
  \endgroup
}

\newcommand{\braket}[2] { \langle #1 | #2 \rangle }

\author{Ruchika Mahajan}
\author{Arti Kashyap}
\affiliation[IIT MANDI]
{School of Basic Sciences, Indian Institute of Technology Mandi, Himachal Pradesh 175075, India}

\author{Iurii Timrov}
\email{iurii.timrov@epfl.ch}
\affiliation[EPFL]
{Theory and Simulation of Materials (THEOS), and National Centre for Computational Design and Discovery of Novel Materials (MARVEL), \'Ecole Polytechnique F\'ed\'erale de Lausanne (EPFL), CH-1015 Lausanne, Switzerland.}

\title{Pivotal Role of Intersite Hubbard Interactions in Fe-doped $\alpha$-MnO$_2$}

\begin{document}






\begin{abstract}
  We present a first-principles investigation of the structural, electronic, and magnetic properties of the pristine and Fe-doped $\alpha$-MnO$_2$ using density-functional theory with extended Hubbard functionals. The onsite $U$ and intersite $V$ Hubbard parameters are determined from first principles and self-consistently using density-functional perturbation theory in the basis of L\"owdin-orthogonalized atomic orbitals. For the pristine $\alpha$-MnO$_2$ we find that the so-called C2-AFM spin configuration is the most energetically favorable, in agreement with the experimentally observed antiferromagnetic ground state. For the Fe-doped $\alpha$-MnO$_2$ two types of doping are considered: Fe insertion in the $2 \times 2$ tunnels and partial substitution of Fe for Mn. We find that the interstitial doping preserves the C2-AFM spin configuration of the host lattice only when both onsite $U$ and intersite $V$ Hubbard corrections are included, while for the substitutional doping the onsite Hubbard $U$ correction alone is able to preserve the C2-AFM spin configuration of the host lattice. The oxidation state of Fe is found to be $+2$ and $+4$ in the case of the interstitial and substitutional doping, respectively, while the oxidation state of Mn is $+4$ in both cases. This work paves the way for accurate studies of other MnO$_2$ polymorphs and complex transition-metal compounds when the localization of $3d$ electrons occurs in the presence of strong covalent interactions with ligands.
\end{abstract}

\section{Introduction}
\label{sec:intro}

\begin{figure}[t]
  \includegraphics[width=0.80\linewidth]{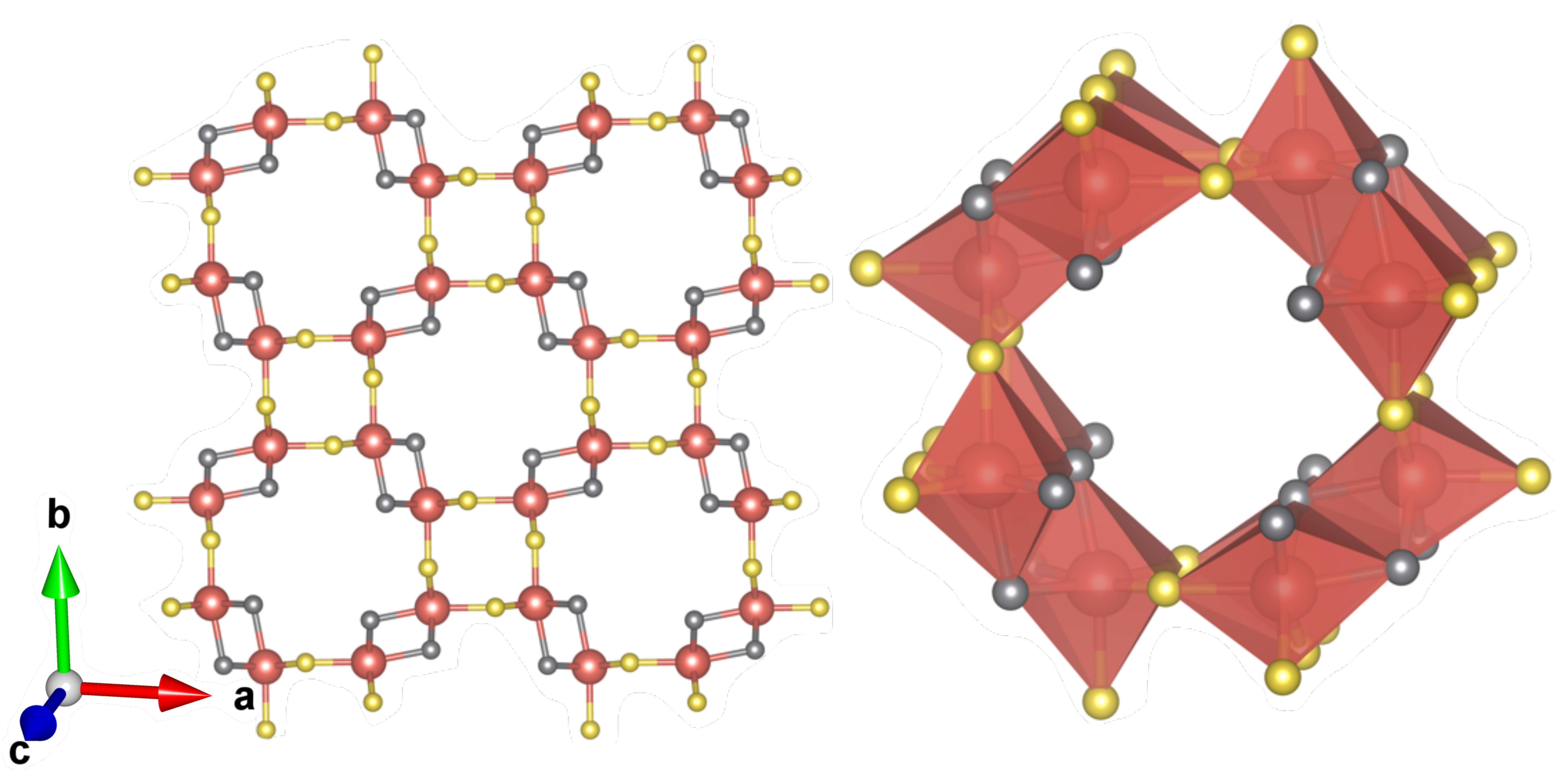}
  \caption{Experimental crystal structure of $\alpha$-MnO$_2$.~\cite{barudvzija2016structural} Mn atoms are indicated in light brown color, while O atoms are indicated in yellow and gray to highlight two types of hybridizations, $sp^2$ and $sp^3$, respectively. Rendered using \textsc{VESTA}~\cite{Momma:2008}.}
\label{fig:crystal_structure}
\end{figure}

Manganese dioxide is an inexpensive and highly reactive material that has attracted great interest due to its structural diversity with different chemical and physical properties.~\cite{Ghosh:2020} In particular, $\alpha$-MnO$_2$ is made up of $2\times 2$ channels due to edge and corner sharing of MnO$_6$ octahedra along the $c$ axis, and there are two types of oxygen atoms having $sp^2$ or $sp^3$ hybridization (see Fig.~\ref{fig:crystal_structure}). Such a tunnel-type structure generally can incorporate external cations (e.g. Li$^+$, K$^+$, NH$^{4+}$, Ba$^{2+}$, Na$^{+}$, Pb$^{2+}$),~\cite{thackeray1997manganese,barudvzija2016structural} transition-metal elements (e.g. Ti, V, Cr, Fe, Co, Ni, Cu, Nb, Ru, Ag),~\cite{brady2019transition, lubke2018transition, hashem2012structural, xu2019effect, song2018first, li2016band,ochoa2016magnetostructural} rare-earth elements (e.g. Ce),~\cite{chen2016phase} and H$_2$O molecules which can substantially alter its properties. $\alpha$-MnO$_2$ has very broad applications, e.g. it is used in Li-, Na-, and Mg-ion batteries,~\cite{zhang2012alpha, debart2008alpha, tompsett2013electrochemistry} Zn-air batteries, supercapacitors, electrochemical energy storage systems~\cite{wu2010high}, catalysts in water oxidation,~\cite{kruthika2017tuning} and many more.~\cite{boppana2011nanostructured, liang2008effect} Experimentally, it is reported to be a semiconductor~\cite{liu2018near} with an antiferromagnetic ordering below $T_N = 24.5$~K, and it crystallizes in a tetragonal crystal structure with a space group $I4/m$ (87).~\cite{yamamoto1974single} Another experimental study showed that spin glass transition occurs in $\alpha$-MnO$_2$ at 50~K due to frustration.~\cite{zhao2012hydrothermal, kumar2017experimental, zhou2018magnetic}

Many studies seek doping in $\alpha$-MnO$_2$ because it has large cavities. The doping enhances the stability of this material~\cite{thackeray1997manganese, barudvzija2016structural, brady2019transition, lubke2018transition, hashem2012structural, xu2019effect, song2018first, li2016band,ochoa2016magnetostructural} and thus it allows for potential applications in the range from catalysts to energy storage systems. Generally, two types of doping are considered: $i)$~the \textit{interstitial} one, where the dopant is placed in the center of the $2 \times 2$ tunnels, and $ii)$~the \textit{substitutional} one, where the dopant is partially substituted for Mn atoms.~\cite{hui2012influence} Of particular interest is the Fe-doping of $\alpha$-MnO$_2$ because Fe atoms have an atomic radius similar to the one of Mn atoms, and hence Fe can partially substitute Mn or it can easily fit into the cavity. $\alpha$-MnO$_2$ doped with Fe is used in industrial processes of preparing cyclohexanone by photocatalytic oxidation process of cyclohexane.~\cite{said2018photo} This compound is not only cheaper but also it shows highest photo-catalytic activity compared to other catalysts.~\cite{said2018photo} Moreover, Fe doping of $\alpha$-MnO$_2$ makes it better catalyst for oxygen evolution reaction and oxygen reduction reaction and an efficient electrode for Zn–air batteries.~\cite{mathur2019one} However, despite numerous experimental studies of the Fe doping of $\alpha$-MnO$_2$~\cite{duan2012controlled, li2017facile, song2022fe, hui2012influence, fan2019fe} the understanding of the changes in the crystal structure, electronic and magnetic properties, and the oxidation state of Fe and Mn ions remains illusive. Only a few theoretical studies have attempted to shed more light on this problem~\cite{song2022fe, duan2013theoretical} since the computational modeling of the pristine and Fe-doped $\alpha$-MnO$_2$ is very challenging due to a complex interplay between structural, electronic, and magnetic degrees of freedom.

First-principles simulations of the pristine and Fe-doped $\alpha$-MnO$_2$ using density-functional theory (DFT)~\cite{Hohenberg:1964,Kohn:1965} with local and semi-local exchange-correlation (xc) functionals encounter serious problems due to the presence of localized and partially-filled $3d$ electrons. In particular, local spin-density approximation (LSDA) and spin-polarized generalized-gradient approximation ($\sigma$-GGA) are unable to produce accurate results due to large self-interaction errors for $3d$ electrons.~\cite{Perdew:1981, mori2006many} Therefore, more accurate approaches that go beyond standard DFT are needed. Many methods have been used to study $\alpha$-MnO$_2$ and most of them are based on Hubbard-corrected DFT (DFT+$U$~\cite{anisimov1991band, liechtenstein1995density, dudarev1998electron})~\cite{tompsett2013electrochemistry, cockayne2012first, crespo2013electronic, kitchaev2016energetics, chepkoech2018first} and DFT with hybrid functionals (e.g. HSE06~\cite{Heyd:2003, Heyd:2006}).~\cite{crespo2013electronic, kitchaev2016energetics} While for hybrid functionals the main bottlenecks are the high computational cost and difficulty to tune the amount of exact exchange, in DFT+$U$ the value of the Hubbard $U$ parameter is unknown and often it is chosen (semi-)empirically which possess ambiguities on the final results of interest. Alarmingly, the calibration procedure of $U$ is highly arbitrary, which often results in incorrect interpretations of experiments, contradictory predictions, and generates a large spread in reported results. 

Most DFT-based studies of $\alpha$-MnO$_2$ consider collinear spin configurations: the ferromagnetic (FM) and various types of the antiferromagnetic (AFM) orderings. In the AFM case, magnetic structures such as A2-AFM, C-AFM, C2-AFM, and G-AFM have been studied (see Fig.~\ref{fig:MagneticStructure}).~\cite{crespo2013electronic, kruthika2017tuning, cockayne2012first, noda2016momentum, kaltak2017charge, ochoa2016magnetostructural} On the one hand, DFT with hybrid functionals predicts FM to be lower in energy than all types of the AFM configurations,~\cite{crespo2013electronic} thus contradicting to the experimental observation of the AFM state.~\cite{yamamoto1974single} On the other hand, DFT+$U$ predicts different trends depending on the value of the Hubbard $U$ parameter. In the vast majority of the DFT+$U$ studies of $\alpha$-MnO$_2$, the $U$ parameter is chosen empirically in the range from 1 to 6~eV ~\cite{crespo2013electronic, cockayne2012first,noda2016momentum} such that DFT+$U$ reproduces well some experimental property of interest (e.g., band gaps, magnetic moments, etc.). However, when selecting $U$ empirically, it is often forgotten to pay attention also to the type of the Hubbard projectors that are used in various electronic-structure codes, as indeed the DFT+$U$ results are extremely sensitive not only to the value of $U$ but also to the type of these projectors.~\cite{wang2016local, Kulik:2008, mahajan2021importance} Only in one DFT+$U$ study of $\alpha$-MnO$_2$ so far the value of $U$ was computed from first principles (using the constrained DFT),~\cite{tompsett2013electrochemistry} though no information is provided regarding the magnetic ordering used. Therefore, a consistent comparison of the FM and various types of AFM orderings in the pristine and Fe-doped $\alpha$-MnO$_2$ using first-principles Hubbard parameters have not been performed so far. Moreover, in all previous Hubbard-corrected DFT studies only the onsite Hubbard $U$ correction was included, while the intersite Hubbard interactions were fully disregarded even though they are known to be very important in MnO$_2$ polymorphs due to strong Mn($3d$)--O($2p$) hybridization, as has been amply demonstrated in Ref.~\cite{mahajan2021importance}.

Herein, we present a fully first-principles study of the structural, electronic, and magnetic properties of the pristine and Fe-doped $\alpha$-MnO$_2$ using DFT with extended Hubbard functionals (DFT+$U$+$V$).~\cite{campo2010extended} The DFT+$U$+$V$ approach has proven to be effective for accurate description of various properties of solids and molecules.~\cite{Kulik:2011, cococcioni2019energetics, Ricca:2020, Timrov:2020c, Cococcioni:2021, mahajan2021importance, Yang:2021, Jang:2022, Yang:2022, Timrov:2022b} We compute the onsite $U$ and intersite $V$ Hubbard parameters using density-functional perturbation theory (DFPT)~\cite{timrov2018hubbard, Timrov:2021} in the basis of L\"owdin-orthogonalized atomic orbitals. Thus, we avoid any empiricism and ambiguities that are so common to the vast majority of other DFT+$U$ studies. The Hubbard parameters are determined using the self-consistent procedure~\cite{hsu2009first, cococcioni2019energetics, Timrov:2021} to ensure the full consistency between the crystal and electronic structures. We find that C2-AFM is the most energetically favorable magnetic ordering in the pristine $\alpha$-MnO$_2$ which agrees with the Goodenough-Kanamori rules.~\cite{goodenough1955theory, kanamori1959superexchange} Overall, the computed crystal structure parameters, band gaps, and magnetic moments are in good agreement with experiments. We find that the interstitial doping preserves the C2-AFM spin configuration of the host lattice only when both onsite $U$ and intersite $V$ Hubbard corrections are included, while for the substitutional doping the onsite Hubbard $U$ correction alone is able to preserve the C2-AFM spin configuration of the host lattice. Finally, using the method of Ref.~\cite{Sit:2011} we find that the oxidation state (OS) of Fe is $+2$ or $+4$ in the case of the interstitial and substitutional doping, respectively, while the OS of Mn is $+4$ irrespective of the doping type. These findings constitute a robust and reliable fully-first-principles characterization of the structural, electronic, and magnetic properties of the pristine and Fe-doped $\alpha$-MnO$_2$ which can serve as a basis for further more advanced studies of this material.

The rest of the paper is organized as follows. Section~\ref{sec:technical_details} contains technical details of the calculations. In Sec.~\ref{sec:pristine_results} we present our findings for the structural, electronic, and magnetic properties of the pristine $\alpha$-MnO$_2$ for different types of collinear magnetic orderings, while Sec.~\ref{sec:Fe_doping} discusses our results for the Fe-doped $\alpha$-MnO$_2$. In Sec.~\ref{sec:Conclusions} we give our conclusions.

\section{Computational details}
\label{sec:technical_details}

\begin{figure}[h]
 \includegraphics[width=0.80\linewidth]{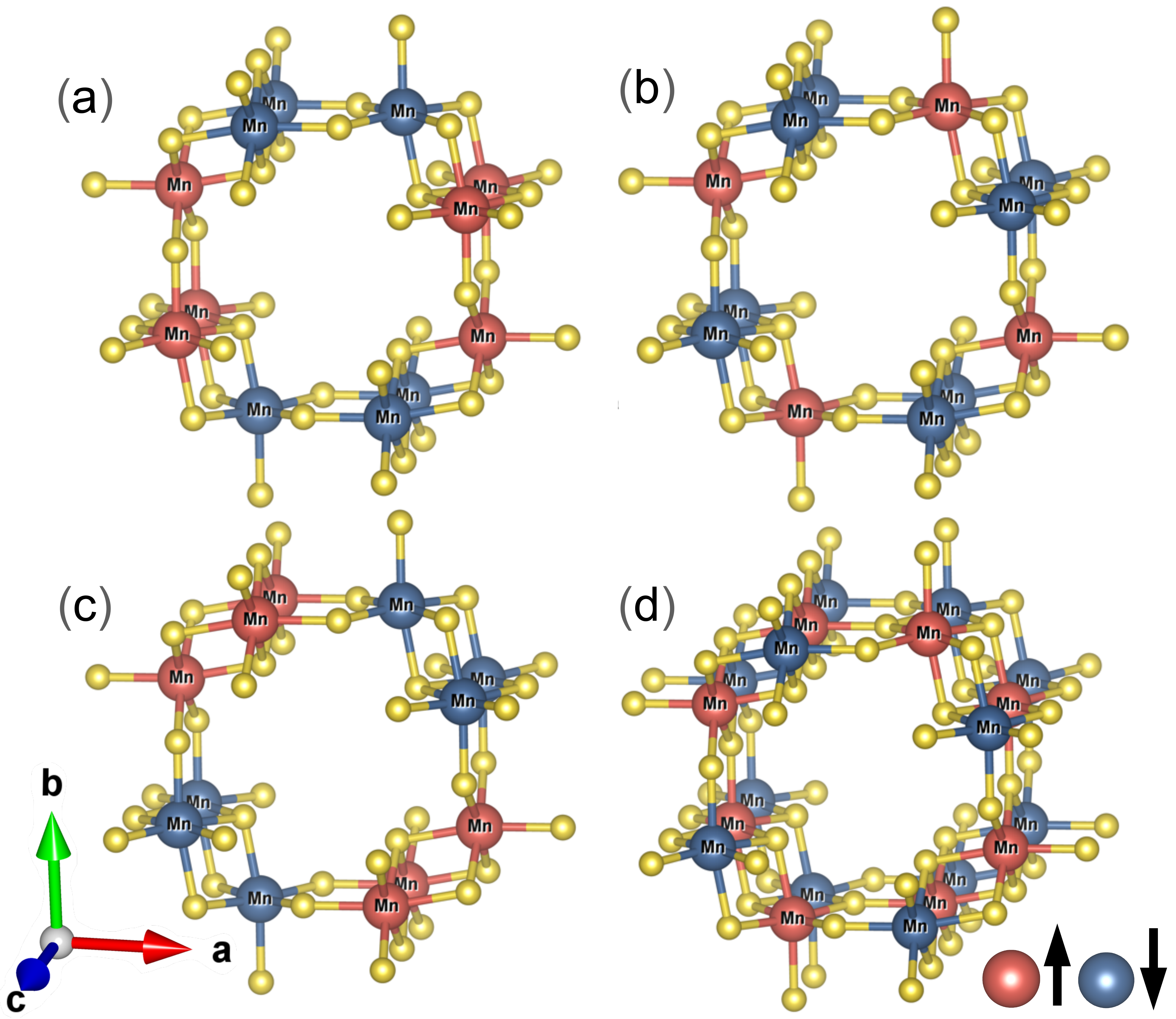}
  \caption{Four collinear AFM configurations of $\alpha$-MnO$_2$: (a)~A2-AFM, (b)~C-AFM, (c)~C2-AFM, and (d)~G-AFM. Mn atoms with spin-up and spin-down alignments are shown in light brown and blue colors, respectively, while oxygen atoms are shown in yellow color. For FM, all Mn atoms have the same spin alignment (not shown). Rendered using \textsc{VESTA}~\cite{Momma:2008}.}
\label{fig:MagneticStructure}
\end{figure}

All calculations were performed using the \textsc{Quantum ESPRESSO} package.~\cite{Giannozzi:2009, Giannozzi:2017, Giannozzi:2020} The computational method is described in Sec.~S1 in supporting information (SI). For the pristine $\alpha$-MnO$_2$, 5 collinear magnetic orderings are considered: FM, A2-AFM, C-AFM, C2-AFM, and G-AFM (see Fig.~\ref{fig:MagneticStructure}). The first 4 magnetic orderings 
have unit cells with 24 atoms, whereas G-AFM was modeled using a supercell of size $1 \times 1 \times 2$ containing 48 atoms. For the Fe-doped $\alpha$-MnO$_2$, also 5 collinear magnetic orderings are considered: interstitial (labeled as ``A'' and ``D'') having 25 atoms and substitutional (``B'', ``C'', and ``E'') having 24 atoms in the simulation cell. 

We have used the xc functional constructed using $\sigma$-GGA with the PBEsol prescription.~\cite{perdew2008restoring} Pseudopotentials were chosen based on the SSSP library~v1.1 (precision):~\cite{prandini2018precision, MaterialsCloud} We have used \texttt{mn\_pbesol\_v1.5.uspp.F.UPF} (GBRV library v1.5~\cite{garrity2014pseudopotentials}), \texttt{O.pbesol-n-kjpaw\_psl.0.1.UPF} and \texttt{Fe.pbesol-spn-kjpaw\_psl.0.2.1.UPF} (Pslibrary v0.3.1~\cite{Kucukbenli:2014}).
For metallic ground states, we used the Marzari-Vanderbilt (MV) smearing~\cite{marzari1999thermal} with a broadening parameter of $2 \times 10^{-2}$~Ry. 
The crystal structure for all spin configurations was optimized at three levels of theory (DFT, DFT+$U$, and DFT+$U$+$V$) using the Broyden-Fletcher-Goldfarb-Shanno (BFGS) algorithm~\cite{Fletcher:1987} with the convergence thresholds of $10^{-6}$~Ry, $10^{-5}$~Ry/Bohr, and $0.5$~KBar for the total energy, forces, and pressure, respectively.
For structural optimizations, the $\mathbf{k}$ points sampling of the first Brillouin zone was done using a uniform $\Gamma$-centered mesh of size $4 \times 4 \times 12$ for all spin configurations except G-AFM for which a $4 \times 4 \times 6$ mesh was used. Kohn-Sham wavefunctions and potentials were expanded in plane waves up to a kinetic-energy cutoff of 90 and 1080~Ry, respectively. Total energies, magnetic moments, band gaps, and the projected density of states (PDOS) were computed using denser $\mathbf{k}$ points meshes of size $8 \times 8 \times 24$ and $8 \times 8 \times 12$, respectively, and the Kohn-Sham wavefunctions and potentials were expanded using higher kinetic-energy cutoffs of 150 and 1800~Ry, respectively (to ensure a $\sim 1$~meV accuracy when comparing the total energies). PDOS was plotted using a Gaussian smearing with a broadening parameter of $4.4 \times 10^{-3}$~Ry.

DFT+$U$ and DFT+$U$+$V$ calculations were performed using the L\"owdin-orthogonalized atomic orbitals as Hubbard projector functions.~\cite{lowdin1950non, mayer2002lowdin, Timrov:2020b, Timrov:2020bb} Hubbard $U$ and $V$ parameters were computed using DFPT~\cite{timrov2018hubbard, Timrov:2021} as implemented in the \textsc{HP} code~\cite{Timrov:2022} which is part of \textsc{Quantum ESPRESSO}. Hubbard parameters will be discussed in detail in the following. We have used uniform $\Gamma$-centered $\mathbf{k}$ and $\mathbf{q}$ points meshes of size $2 \times 2 \times 6$ and $1 \times 1 \times 3$, respectively, for all spin configurations (for the pristine and Fe-doped $\alpha$-MnO$_2$) except G-AFM for which we have used $\mathbf{k}$ and $\mathbf{q}$ points meshes of size $2 \times 2 \times 4$ and $1 \times 1 \times 2$, respectively. Kohn-Sham wavefunctions and potentials were expanded in plane waves up to a kinetic-energy cutoff of 60 and 720~Ry, respectively, giving the accuracy of computed Hubbard parameters of $\sim 0.01$~eV. It is important to stress that we have used a self-consistent procedure for the calculation of $U$ and $V$ as described in detail in Ref.~\cite{Timrov:2021} which consists of cyclic calculations containing structural optimizations and recalculations of Hubbard parameters for each new geometry.

The data used to produce the results of this paper are available in the Materials Cloud Archive.~\cite{MaterialsCloudArchive2022}

\section{Results and Discussion}
\label{sec:Results_and_Discussion}

\subsection{Pristine $\alpha$-MnO$_2$}
\label{sec:pristine_results}

In this section we present our findings for the pristine $\alpha$-MnO$_2$. First we discuss the computed Hubbard parameters, and then we analyze the structural, magnetic, and electronic properties of this material.

\subsubsection{Hubbard parameters}
\label{sec:Hubbard_parameters}

In previous studies it was shown that the values of Hubbard parameters depend on a type of Hubbard projectors, pseudopotentials, oxidation state of transition-metal elements, xc functional, chemical environment, etc.~\cite{Kulik:2008, nawa2018scaled, kick2019intricacies, kirchner2021extensive, mahajan2021importance} In particular, in Ref.~\cite{mahajan2021importance} it is shown for $\beta$-MnO$_2$ that the values of Hubbard $U$ parameters for Mn$(3d)$ states differ by $1.4 - 2.2$~eV depending on whether nonorthogonalized or orthogonalized atomic orbitals are used as Hubbard projector functions. Since it was concluded that the highest accuracy is achieved when using orthogonalized atomic orbitals, here we present results obtained using this type of Hubbard projectors.

Table~\ref{tab:Hub_param} shows self-consistent Hubbard parameters for five collinear magnetic configurations of $\alpha$-MnO$_2$ (FM, A2-AFM, C-AFM, C2-AFM, and G-AFM) computed using DFPT in the basis of L\"owdin-orthogonalized atomic orbitals as described in Sec.~S1 in SI. As in the case of $\beta$-MnO$_2$, for $\alpha$-MnO$_2$ we find that the value of Hubbard $U$ varies marginally depending on the type of the magnetic ordering. When using the extended formulation of DFT+$U$ by incorporating also intersite $V$, the value of $U$ increases slightly due to changes in the electronic screening. In practice, in the DFT+$U$+$V$ framework the size of the response matrices $\chi$ and $\chi_0$ is larger than in the DFT+$U$ case because we compute the response not only of Mn($3d$) states but also of O($2p$) states. Hence, when inverting these response matrices there is a mixing of all matrix elements including those that describe a coupling of Mn($3d$) and O($2p$) states, and the final values of $U$ and $V$ are obtained according to Eq.~(S4) in SI.

As can be seen in Table~\ref{tab:Hub_param}, the intersite Hubbard $V$ parameters are smaller than the onsite Hubbard $U$ by a factor of $6-9$. Nevertheless, the relatively small $V$ values turn out to be important to improve the accuracy of various ground-state properties of $\alpha$-MnO$_2$ as will be shown in the following, in analogy with findings for $\beta$-MnO$_2$.~\cite{mahajan2021importance}

\begin{table*}[t]
\renewcommand{\arraystretch}{1.2}
\centering
\begin{tabular}{c|c|ccccc}
\hline
\multirow{2}{*}{Method}      & \multirow{2}{*}{HP}   & \multicolumn{5}{c}{Magnetic ordering}           \\ \cline{3-7} 
                             &     &    FM            & A2-AFM               & C-AFM                 & C2-AFM                 & G-AFM   \\ \hline
\multirow{1}{*}{DFT+$U$}     & $U$ & $6.67$           & $6.53$               & $6.41$                & $6.47$                 & $6.43$        \\ \hline
\multirow{2}{*}{DFT+$U$+$V$} & $U$ & $6.83$           & $6.74$               & $6.65$                & $6.67$                 & $6.73$        \\ 
                             & $V$ & $0.72-1.09$      & $0.75-1.07$          & $0.71-1.12$           & $0.66-1.13$            & $0.76-1.16$   \\ 
\hline
\end{tabular}%
\caption{Self-consistent Hubbard parameters (HP) in eV computed using DFPT for five magnetic orderings of the pristine $\alpha$-MnO$_2$: FM, A2-AFM, C-AFM, C2-AFM, and G-AFM. The onsite $U$ for Mn($3d$) states and intersite $V$ between Mn($3d$) and O($2p$) states are computed in the frameworks of DFT+$U$ and DFT+$U$+$V$ (PBEsol functional) using L\"owdin-orthogonalized atomic orbitals as Hubbard projector functions.}
\label{tab:Hub_param}
\end{table*}

\subsubsection{Structural properties}
\label{sec:Structural_properties}

$\alpha$-MnO$_2$ has a tetragonal crystal structure which is described by two lattice parameters $a$ and $c$.~\cite{yamamoto1974single} The detailed comparison of these lattice parameters and the cell volume $V$ at three levels of theory and for different magnetic orderings is presented in Table~S1 in SI, while here we discuss briefly only the general trends. DFT underestimates the $a$ parameter, while $c$ is very close to the experimental value, and hence overall the cell volume is also underestimated by $3-4$~\%. In contrast, adding the Hubbard $U$ correction leads to $a$ and $c$ which are larger than the experimental ones, and the cell volume is overestimated by $3-6$~\%. The best agreement with the experiments is achieved at the level of DFT+$U$+$V$: the optimized $a$ parameter falls in the range of experimental values or slightly overestimated, while $c$ is also slightly overestimated, and the cell volume is within $1-3$~\% compared to experiments. Within DFT+$U$+$V$, three types of magnetic orderings (C-AFM, C2-AFM, and G-AFM) give the most accurate predictions of the lattice parameters and the cell volume. Importantly, at all levels of theory we find that the cell remains tetragonal in agreement with experiments and in contrast to previous works that reported a small monoclinic distortion.~\cite{cockayne2012first, crespo2013electronic} Such a discrepancy with respect to previous theoretical studies could be due to the fact that in this work we use more stringent computational setup when performing structural optimizations. 

To be more specific, we focus on the C2-AFM ordering that was also studied in detail in previous works~\cite{cockayne2012first, crespo2013electronic} and because this is the most energetically favorable ordering as will be shown in Sec.~\ref{sec:Energetics}. Table~\ref{tab:C2AFMBondLengthAngles} summarizes the crystal structure parameters, bond lengths and angles for this spin configuration that were optimized at the levels of DFT, DFT+$U$, and DFT+$U$+$V$ and as measured in experiments. It can be seen that the most accurate agreement with experiments is achieved using DFT+$U$+$V$. In particular, when using DFT+$U$+$V$ we find two values of bond lengths consistently with experiments, while when using DFT and DFT+$U$ there are some variations in bond lengths due to lowering of the symmetry. These findings underline the importance not only of the onsite Hubbard corrections but also of the intersite ones when optimizing the crystal structure and atomic positions for the pristine $\alpha$-MnO$_2$.  
In the following sections we present various ground-state properties of the pristine $\alpha$-MnO$_2$ computed at different levels of theory and for different magnetic orderings using respective crystal structure parameters.

\begin{table*}[t]
\renewcommand{\arraystretch}{1.3}
\centering
\begin{tabular}{ccccc}
\hline
 CSP       & DFT        & DFT+$U$     & DFT+$U$+$V$ & Expt. \\
\hline 
$a$        & 9.66       & 9.85        &  9.78       & 9.75$^a$, 9.79$^c$, 9.84$^b$ \\
$c$        & 2.83       & 2.93        &  2.91       & 2.86$^{a, b}$, 2.87$^{c}$ \\
$V$        & 263.9      & 284.6       &  278.3      & 272.0$^a$, 274.5$^{c}$, 276.9$^b$ \\
$d_1$      & 1.85, 1.87 & 1.90, 1.92  &  1.90       & 1.89$^d$ \\
$d_2$      & 1.88, 1.91 & 1.94        &  1.93       & 1.93$^d$ \\
$\theta_1$ & 131.0      & 129.5       &  129.8      & 130.2$^e$ \\
$\theta_2$ & 98.0       & 98.6        &  98.3       & 98.2$^e$ \\
\hline
\end{tabular}
\vskip 0.1cm
\hspace{-0.2cm} $^a$Ref.~\cite{thackeray1997manganese}
\hspace{0.2cm}  $^b$Ref.~\cite{islam2017carbon}
\hspace{0.2cm}  $^c$Ref.~\cite{rossouw1992alpha}
\hspace{0.2cm}  $^d$Ref.~\cite{robinson2013photochemical}
\hspace{0.2cm}  $^e$Refs.~\cite{mathur2019one} 
\caption{Comparison of experimental and theoretical (computed for C2-AFM using DFT, DFT+$U$, and DFT+$U$+$V$) crystal structure parameters (CSP) for the pristine $\alpha$-MnO$_2$ [see Fig.~\ref{fig:MagneticStructure}~(c)]: the lattice parameters $a$ and $c$ (in \AA), the cell volume $V$ (in \AA$^3$), the bond lengths $d_1$ and $d_2$ (in \AA), and the bond angles $\theta_1$ and $\theta_2$. DFT and DFT+$U$ structures have lower symmetry, hence the bond lengths $d_1$ (and $d_2$) take multiple values.}
\label{tab:C2AFMBondLengthAngles}
\end{table*}

\subsubsection{Energetics}
\label{sec:Energetics}

\begin{figure}[t]
 \includegraphics[width=0.80\linewidth]{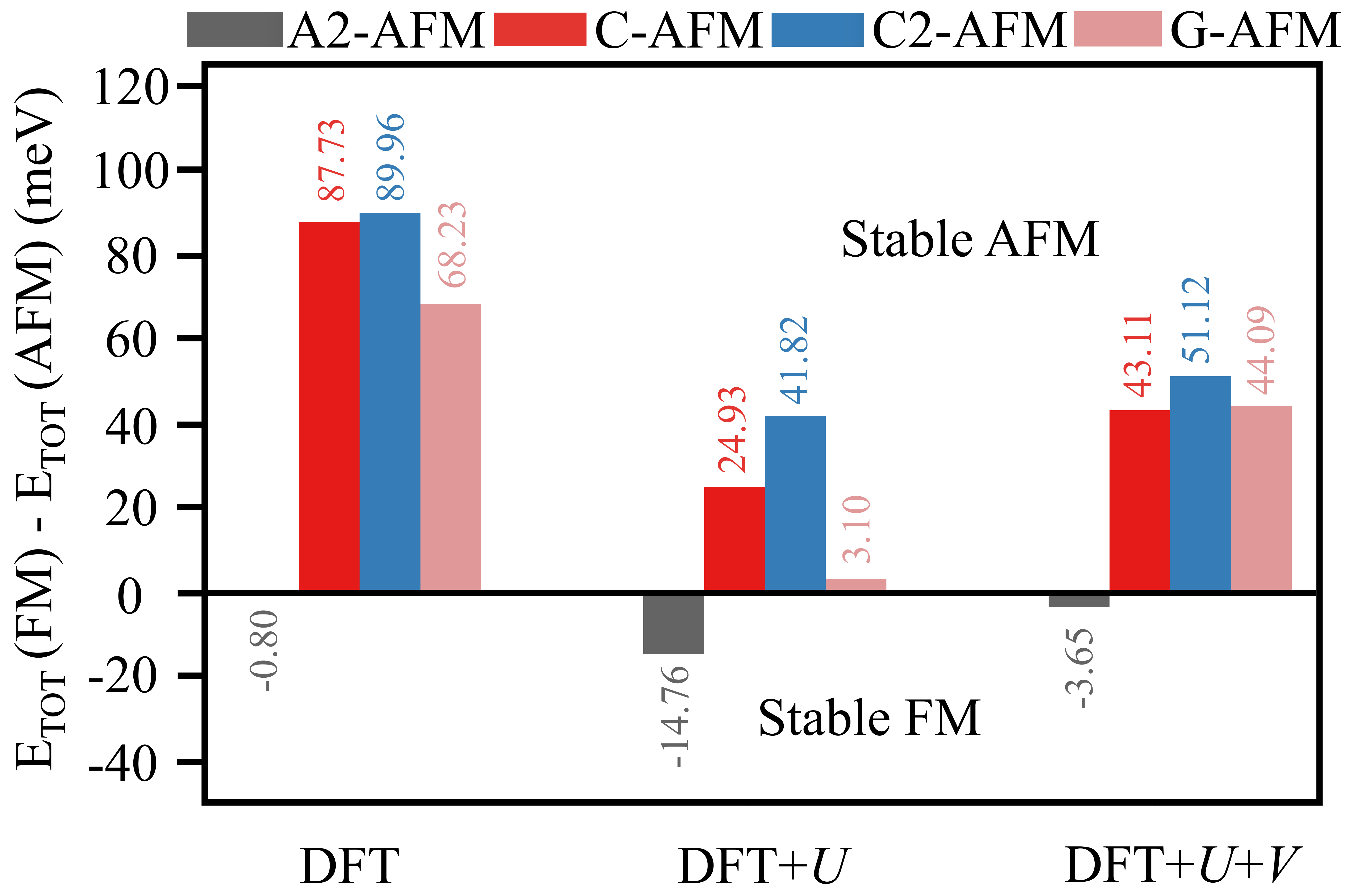} 
 \caption{Total energy difference per formula unit (in~meV) for five collinear magnetic orderings of the pristine $\alpha$-MnO$_2$ (FM, A2-AFM, C-AFM, C2-AFM, G-AFM) computed at three levels of theory (DFT, DFT+$U$, and DFT+$U$+$V$) using the PBEsol functional. For each case, the Hubbard parameters $U$ and $V$ were computed using DFPT and L\"owdin-orthogonalized atomic orbitals, and they are listed in Table~\ref{tab:Hub_param}. Positive values on the plot mean that the AFM ordering is lower in energy than FM and hence the former is more energetically favorable, while negative values mean the opposite. The largest positive value on the plot (at each level of theory) corresponds to the lowest-energy magnetic ordering.}
\label{fig:RE}
\end{figure}

Experimentally it is known that the pristine $\alpha$-MnO$_2$ has an AFM ordering below the N\'eel temperature of 24.5~K,~\cite{yamamoto1974single} however the precise direction of the atomic magnetic moments is not known. In this respect, computational studies based on DFT could be useful as they might suggest what type of the magnetic ordering is the most energetically favorable. In particular, DFT+$U$ predictions are very sensitive not only to the value of $U$ but also to the type of Hubbard projector functions. For example, in Ref.~\cite{crespo2013electronic} it is reported that for the effective $U \geq 2.0$~eV and when using nonorthogonalized atomic orbitals 
as Hubbard projector functions the FM ordering is the most energetically favorable compared to all types of the AFM ordering. In contrast, in Refs.~\cite{zhou2017spatially, noda2016momentum, kitchaev2016energetics} it was found that FM is the most energetically favorable for the effective $U \gtrsim 4.0$~eV and when using projector augmented wave (PAW) Hubbard projectors. In this work we do not use empirical values of $U$ contrary to previous works, and hence here we do not provide ranges of $U$ for which FM or AFM is the most energetically favorable. Instead, we use first-principles Hubbard parameters reported in Table~\ref{tab:Hub_param} respectively for each spin configuration and check which one has the lowest energy. We use exactly the same approach as in Ref.~\cite{mahajan2021importance} for $\beta$-MnO$_2$ (in particular, see the discussion in Sec.~IV~C of Ref.~\cite{mahajan2021importance}). Interestingly, we find that using our first-principles values of Hubbard parameters ($U \approx 6-7$~eV and $V \approx 0.7-1.2$~eV, see Table~\ref{tab:Hub_param}) and L\"owdin-orthogonalized atomic orbitals, three types of the AFM ordering (C-AFM, C2-AFM, and G-AFM) are more energetically favorable than the FM one (see Fig.~\ref{fig:RE}). This finding clearly shows that the choice of Hubbard projector functions is as crucial as the values of Hubbard parameters, and hence both should be considered when comparing conclusions from different works. However, one should keep in mind that there are also other differences between this work and previous studies, such as e.g. different pseudopotentials, which could also affect the final DFT+$U$(+$V$) predictions.~\cite{Kulik:2008} 

As can be seen in Fig.~\ref{fig:RE}, C2-AFM is the most energetically favorable spin configuration at all levels of theory. However, at the levels of DFT+$U$ and DFT+$U$+$V$, C2-AFM has larger energy difference compared to C-AFM and G-AFM. This shows that Hubbard-corrected DFT stabilizes C2-AFM more than standard DFT. It is noteworthy that DFT+$U$ and DFT+$U$+$V$ convey similar trends for the energetics of the pristine $\alpha$-MnO$_2$, hence the intersite $V$ corrections turn out to be no so crucial contrary to what is the case for $\beta$-MnO$_2$.~\cite{mahajan2021importance} However, it will be shown in the following that for the Fe-doped $\alpha$-MnO$_2$ the intersite Hubbard corrections are crucial to stabilize the C2-AFM spin configuration of the host lattice. Finally, we note that the C2-AFM ordering is consistent with the Goodenough-Kanamori rules,~\cite{goodenough1955theory, kanamori1959superexchange} which state that superexchange interactions favor AFM coupling when the Mn-$3d$ and O-$2p$ orbitals overlap as in a linear (180$^\circ$) Mn$-$O$-$Mn group (point sharing of MnO$_6$ octahedra, see Fig.~\ref{fig:crystal_structure}), while they favor FM coupling when the Mn-$3d$ and O-$2p$ orbitals overlap as in a bent (90$^\circ$) Mn$-$O$-$Mn group (edge sharing of MnO$_6$ octahedra, see Fig.~\ref{fig:crystal_structure})~\cite{noda2016momentum}. Even though the bond angles are not 90$^\circ$ and 180$^\circ$ as required by the Goodenough-Kanamori rules (see Table~\ref{tab:C2AFMBondLengthAngles}), we still find that C2-AFM is the most energetically favorable magnetic ordering.

\subsubsection{Magnetic moment}
\label{sec:magnetic_moments}

\begin{figure}[h]
 \includegraphics[width=0.80\linewidth]{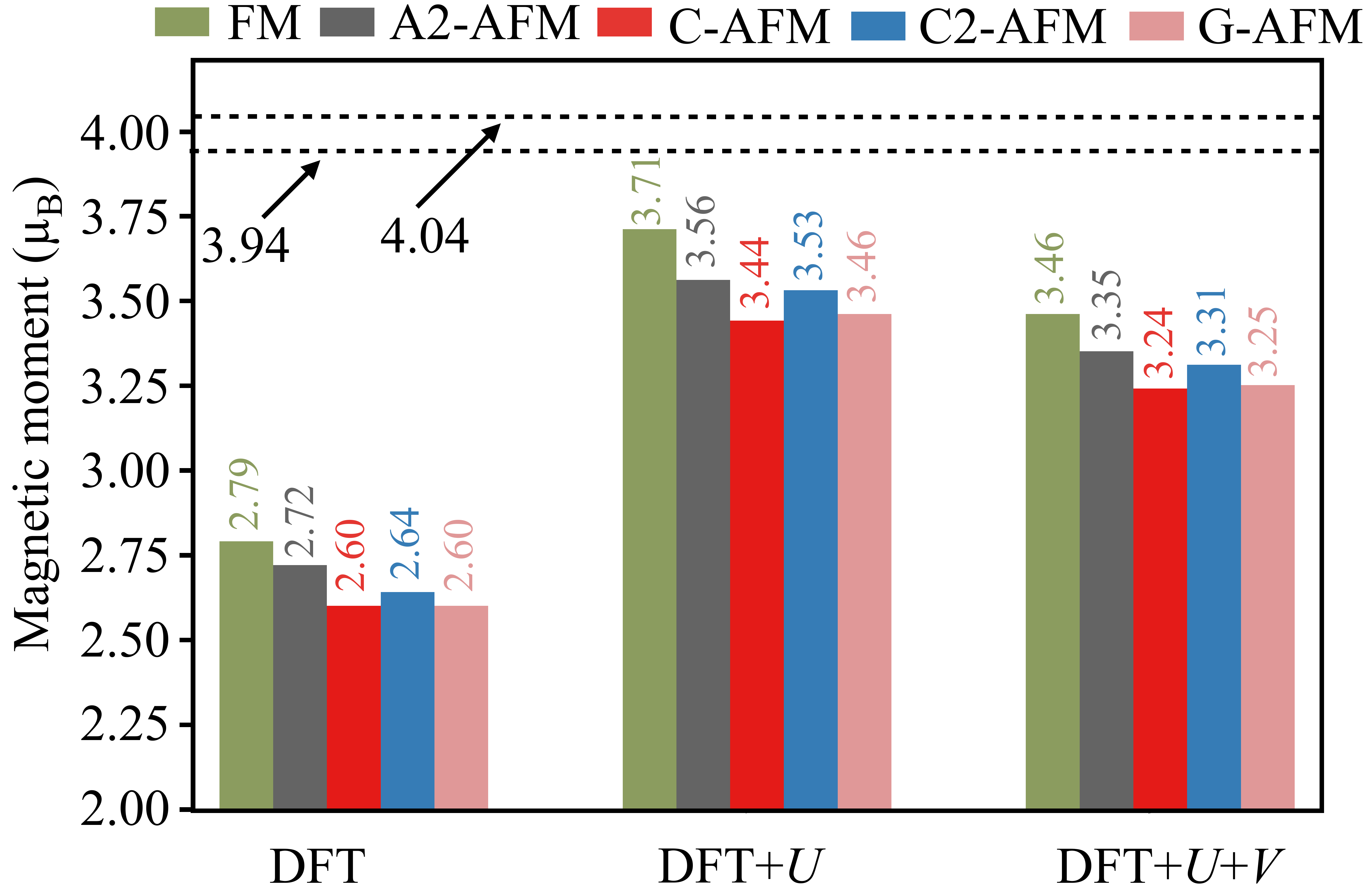} 
\caption{Magnetic moment (in $\mu_\mathrm{B}$) on Mn atoms in the pristine $\alpha$-MnO$_2$ for five collinear magnetic orderings (FM, A2-AFM, C-AFM, C2-AFM, G-AFM) computed at three levels of theory (DFT, DFT+$U$, and DFT+$U$+$V$) using the PBEsol functional. For each case, the Hubbard parameters $U$ and $V$ were computed using DFPT and are listed in Table~\ref{tab:Hub_param}. The experimental magnetic moments are $3.94-4.04$~$\mu_\mathrm{B}$~\cite{li2007one} and $4.04$~$\mu_\mathrm{B}$~\cite{kumar2017experimental} and they are indicated with horizontal dashed lines.}
\label{fig:Magnetic_moment}
\end{figure}

Figure~\ref{fig:Magnetic_moment} shows a comparison of the magnetic moments on Mn atoms in the pristine $\alpha$-MnO$_2$ as measured in experiments and as computed using three levels of theory for five magnetic orderings. The magnetic moments were determined using the projection method, i.e. by computing the onsite occupation matrix ($I=J$) using Eq.~(S3) in SI and then by taking a difference between the spin-up and spin-down components. It can be seen in Fig.~\ref{fig:Magnetic_moment} that DFT underestimates the magnetic moments by about $29-36$\% depending on the type of the magnetic ordering. DFT+$U$ provides the closest agreement with experiments with magnetic moments being underestimated by $6-15$\%, while DFT+$U$+$V$ gives magnetic moments that fall in the range between the DFT and DFT+$U$ predictions. This result is not surprising, since the onsite $U$ correction localizes the $3d$ electrons on the Mn atoms and reduces the hybridization with the ligand states, and thus the magnetic moments are increased. Instead, within DFT+$U$+$V$ the hybridization with ligands is partially restored while still preserving the localized nature of the $3d$ electrons, and as a consequence the magnetic moments are slightly decreased. 

For our best candidate (i.e. C2-AFM) with our best approach (DFT+$U$+$V$) the computed magnetic moment is 3.31~$\mu_\mathrm{B}$ which differs from the experimental ones by $16-18$\%. Thus, there is definitively a room for improvements, and this could be possibly achieved by including also the Hund's $J$ interactions,~\cite{himmetoglu2011first, bajaj2017communication, linscott2018role} but this goes beyond the frames of this study.

\subsubsection{Band gap}
\label{sec:Band_gap}

Pristine $\alpha$-MnO$_2$ is a semiconductor, but the exact value of the band gap is difficult to determine experimentally since it is hard to synthesize pure samples without dopants. Noticeable works are Refs.~\cite{liu2018near, salari2020facile, shah2019study} that reported experimental band gap values in the range from 1.61 to 2.10~eV. Previous computational studies reported quite scattered values of band gaps depending on the method used. For example, in Ref.~\cite{crespo2013electronic} the HSE06 functional was used that predicted FM as the ground state with the respective band gap of 2.2~eV. In the same work, the authors used DFT+$U$ with the empirical $U=1.6$~eV and nonorthogonalized atomic orbitals as Hubbard projectors predicting C2-AFM to be the most energetically stable with a band gap of 0.94~eV. In another DFT+$U$ study,~\cite{cockayne2012first} empirical $U$ = 2.8~eV and $J$ = 1.2~eV with PAW Hubbard projectors were used giving the band gap of 1.33~eV for C2-AFM. Finally, in the DFT+$U$ study of Ref.~\cite{ling2012capture} the empirical $U$ = 3.9~eV with PAW Hubbard projectors was used for C2-AFM giving the gap of 0.70~eV. Hence, in all these studies the band gap values lie outside the experimental range.

\begin{figure}[t]
 \includegraphics[width=0.80\linewidth]{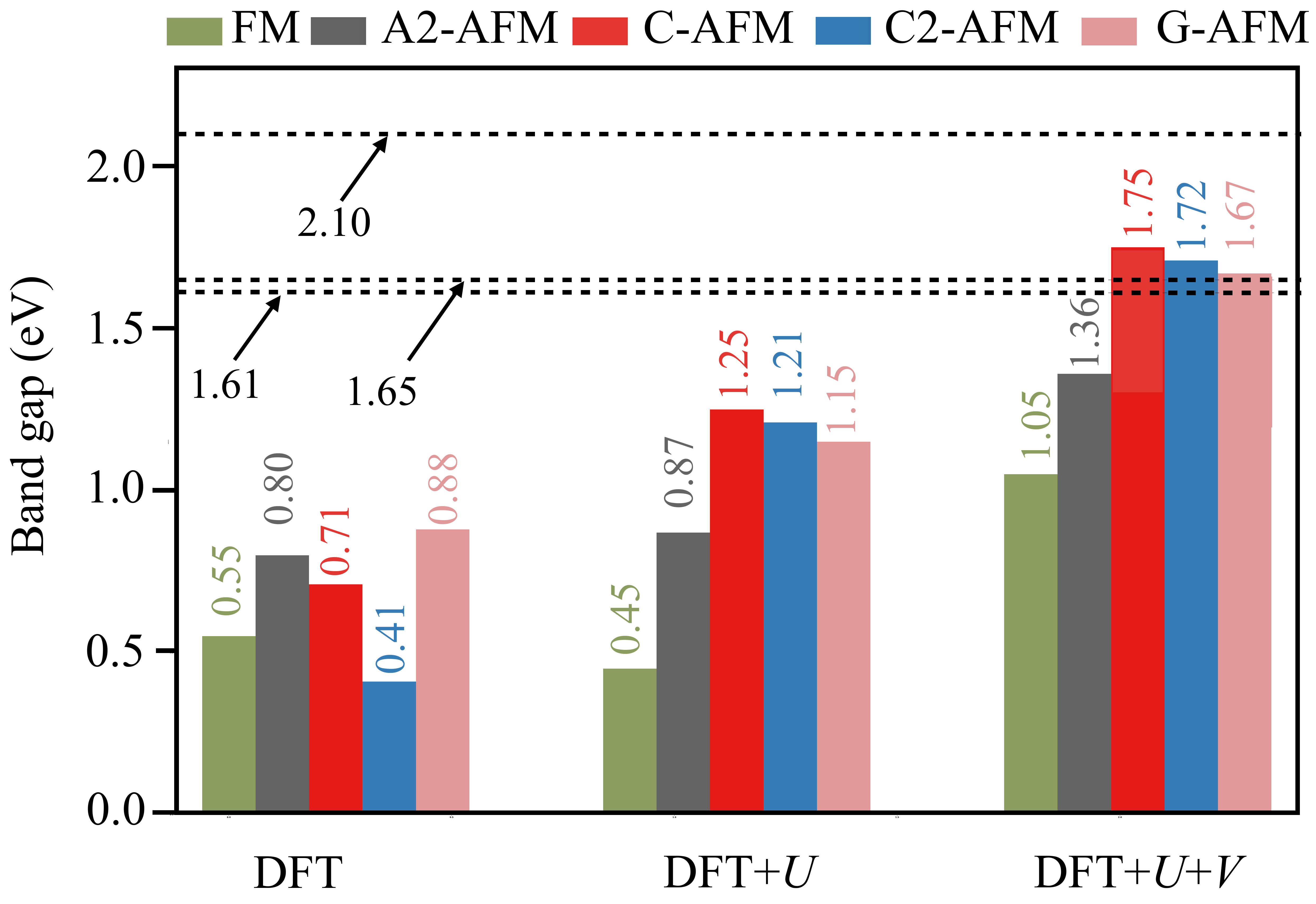}
 \caption{Band gap (in~eV) in the pristine $\alpha$-MnO$_2$ for five collinear magnetic orderings (FM, A2-AFM, C-AFM, C2-AFM, G-AFM) computed at three levels of theory (DFT, DFT+$U$, and DFT+$U$+$V$) using the PBEsol functional. The experimental band gaps (1.61~eV,~\cite{liu2018near} 1.65~eV,~\cite{salari2020facile} and 2.10~eV~\cite{shah2019study}) are indicated with horizontal dashed lines.}
\label{fig:Band_gaps}
\end{figure}

Figure~\ref{fig:Band_gaps} shows a comparison of band gaps computed using DFT, DFT+$U$, and DFT+$U$+$V$ for five magnetic orderings of the pristine $\alpha$-MnO$_2$. We can see that on average DFT underestimates the band gaps the most, followed by DFT+$U$ and then by DFT+$U$+$V$. As in the case of $\beta$-MnO$_2$, the band gap is very sensitive to the type of the magnetic ordering.~\cite{mahajan2021importance} Here we find that DFT+$U$+$V$ again provides the most accurate band gaps, with C-AFM, C2-AFM, and G-AFM having band gaps that fall in the range of the experimental values. In particular, the band gap for the C2-AFM ordering, which is the most energetically favorable, is 1.72~eV which is very close to the lower bound of the known experimental values ($1.61-1.65$~eV). Definitely, the narrowing down of the experimental range of band gaps is needed in order to further assess the predictive accuracy of DFT+$U$+$V$ - this requires further experimental scrutiny.

\subsubsection{Projected density of states}
\label{sec:PDOS}

\begin{figure}[H]
\begin{center}
  \includegraphics[angle=0,width=0.49\textwidth]{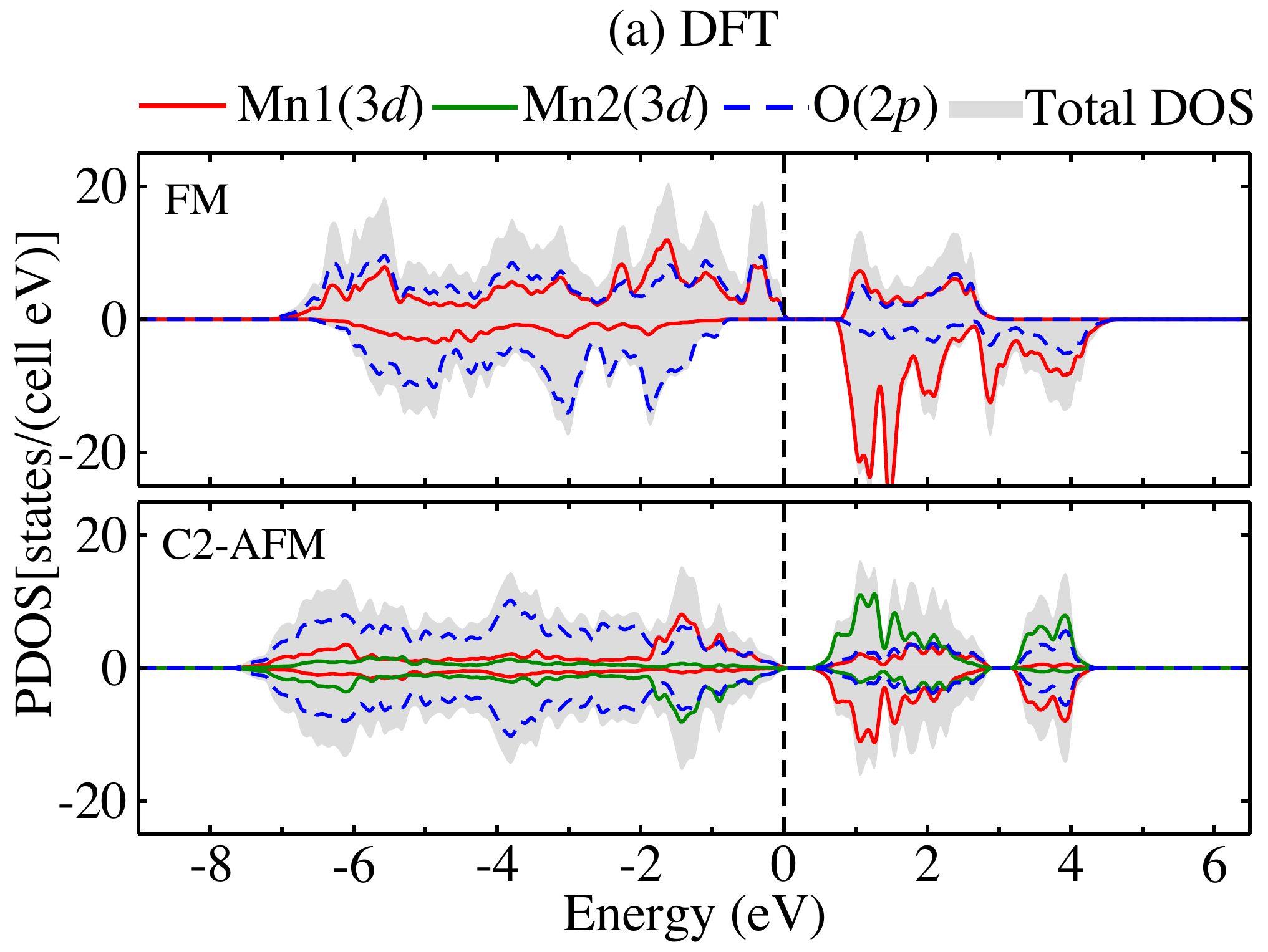}
  \vskip 0.3 cm
  \includegraphics[angle=0,width=0.49\textwidth]{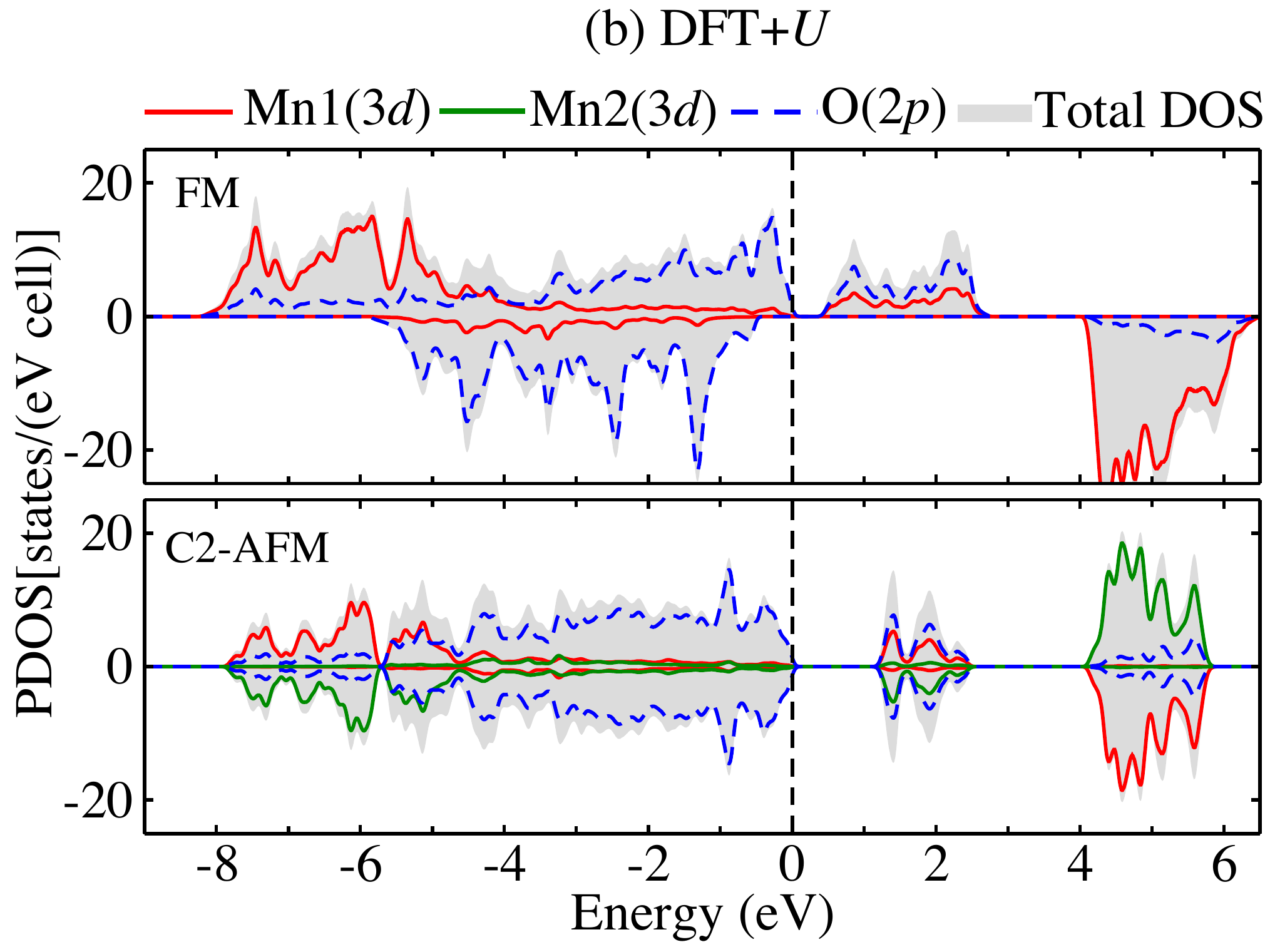}
  \vskip 0.3 cm
  \includegraphics[angle=0,width=0.49\textwidth]{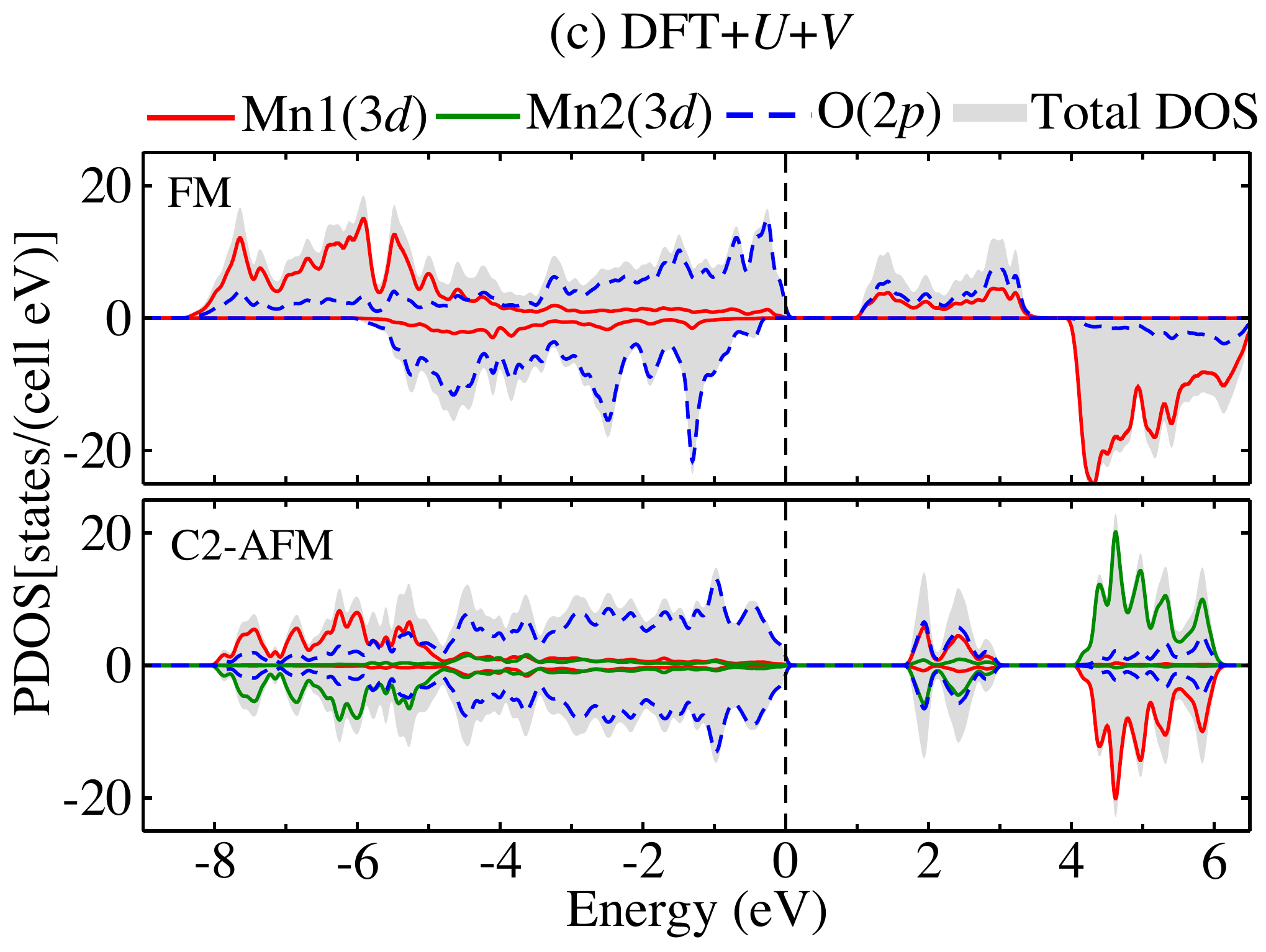}
  \caption{Spin-resolved PDOS and total DOS for the FM and C2-AFM magnetic orderings of the pristine $\alpha$-MnO$_2$ computed at three levels of theory using the PBEsol functional: (a)~DFT, (b)~DFT+$U$, and (c)~DFT+$U$+$V$. For each case, the Hubbard parameters $U$ and $V$ were computed using DFPT using L\"owdin-orthogonalized atomic orbitals and are listed in Table~\ref{tab:Hub_param}. The zero of energy corresponds to the top of the valence bands.}
\label{fig:pdos_pristine}
\end{center}
\end{figure}

Figure~\ref{fig:pdos_pristine} shows the spin-resolved PDOS and total DOS for the FM and C2-AFM magnetic orderings of the pristine $\alpha$-MnO$_2$ computed at three levels of theory, while in Fig.~S1 in SI we show the spin-resolved PDOS for the other three AFM magnetic orderings. The spin-up (upper part) and spin-down (lower part) components of the PDOS are shown on each panel in Fig.~\ref{fig:pdos_pristine} and they were obtained by summing over all atoms of the same type. Mn1 and Mn2 correspond to Mn atoms with the opposite spin polarizations in the AFM case. In general, the trends for $\alpha$-MnO$_2$ are very similar to those that were observed for $\beta$-MnO$_2$.~\cite{mahajan2021importance}

In the DFT case, we see that the Mn($3d$) states are very delocalized due to large self-interaction errors, and they hybridize strongly with the O($2p$) states. The exact shape of the PDOS varies depending on the type of magnetic ordering. However, qualitatively the PDOS for various AFM types are very similar and differ mainly by the magnitude of the band gap and details in the dispersion of the valence and conduction bands. The top of the valence bands is of a mixed Mn($3d$) and O($2p$) character, while the bottom of the conduction bands is dominated by the Mn($3d$) states. 

The PDOS at the DFT+$U$ level of theory differs strongly from the DFT one. As in the case of $\beta$-MnO$_2$, the application of the Hubbard $U$ correction pushes the occupied Mn($3d$) states to lower energies while the empty Mn($3d$) states are pushed to higher energies. As a consequence, the top of the valence bands becomes of a pure O($2p$) character, while the bottom of the conduction bands becomes of a mixed Mn($3d$) and O($2p$) character. As in Ref.~\cite{crespo2013electronic} we find that for FM the empty $e_g$ states in the spin-up channel overlap with the empty $t_{2g}$ and $e_g$ states in the spin-down channel in the DFT case, while in DFT+$U$ these empty states in the spin-down channel are shifted to higher energies and there is an energy gap between these states and the empty $e_g$ states in the spin-up channel. For AFM, the trends are the same and they apply to Mn1 and Mn2 atoms that have opposite spin polarizations. Finally, within DFT+$U$+$V$ the PDOS looks very similar to the one computed using DFT+$U$. One of the major differences is that the band gap is larger in DFT+$U$+$V$ than in DFT+$U$, as discussed in Sec.~\ref{sec:Band_gap} and the energy separation between the groups of empty bands is somewhat smaller in DFT+$U$+$V$.

We are not aware of any valence- or conduction-band spectra measured in the X-ray photoelectron spectroscopy (XPS) and X-ray absorption near-edge structure (XANES) experiments for the pristine $\alpha$-MnO$_2$. Hence, at this stage we cannot verify the accuracy of the computed spin-polarized PDOS. However, these results are valuable for further discussions when considering the Fe-doped $\alpha$-MnO$_2$. In particular, our goal is to understand where the Fe($3d$) states are located and whether the semiconducting character of $\alpha$-MnO$_2$ is preserved after doping. All this is discussed in Sec.~\ref{sec:FedopedPDOS}.

\subsection{Fe-doped $\alpha$-MnO$_2$}
\label{sec:Fe_doping}
\begin{figure*}[h]
 \includegraphics[width=\textwidth,height=3.5cm]{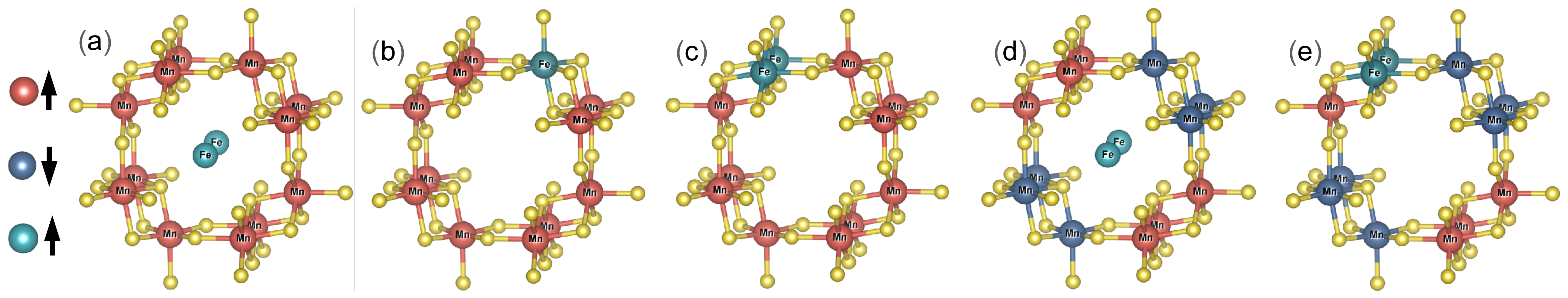}
 \caption{Five collinear spin configurations for the Fe-doped $\alpha$-MnO$_2$: (a)~A, (b)~B, (c)~C, (d)~D, and (e)~E. A and D correspond to the interstitial doping, while B, C, and E correspond to the substitutional doping. Light brown and blue colors correspond to Mn atoms with spin-up and spin-down alignments, respectively, while light green color corresponds to the Fe atom with the spin-up alignment. The oxygen atoms are shown in yellow color. Rendered using \textsc{VESTA}~\cite{Momma:2008}.}
 \label{fig:Fedoped_structures}
\end{figure*}

In this section we discuss structural, electronic, and magnetic properties of the Fe-doped $\alpha$-MnO$_2$. We examine two types of doping, namely the interstitial one (FeMn$_8$O$_{16}$) and the substitutional one (FeMn$_7$O$_{16}$). For the former case we consider two spin configurations (labeled as ``A'' and ``D''), while for the latter we consider three cases (``B'', ``C'', and ``E''). These spin configurations are based on the C2-AFM and FM magnetic orderings of the pristine $\alpha$-MnO$_2$, since the former is the most energetically favorable one while the latter is also often considered in various studies. All these five cases of Fe-doped $\alpha$-MnO$_2$ differ by the variations in the chemical environment and magnetic interactions for Fe atoms and are schematically represented in Fig.~\ref{fig:Fedoped_structures}. We also note that the configurations A, B, and C have the FM ordering, while D and E have the ferrimagnetic (FiM) ordering. 

\subsubsection{Hubbard parameters}
\label{sec:FeStructural_properties}

In the previous DFT+$U$ studies of the Fe-doped $\alpha$-MnO$_2$,~\cite{duan2013theoretical, song2022fe} the value of Hubbard $U$ for Mn($3d$) states was chosen empirically and it was assumed to be exactly the same as in the pristine $\alpha$-MnO$_2$. Here we show that when $U$ is computed from first principles, its value for Mn($3d$) states changes due to the Fe doping, since the latter leads to changes in the crystal and electronic structure. Indeed, from Table~\ref{tab:Fe-dopedHub_param} we can see that on average the $U$ parameter for Mn($3d$) states is increased by a fraction of an eV compared to the pristine $\alpha$-MnO$_2$ case (cf. Table~\ref{tab:Hub_param}). Moreover, there is no longer just one global value of $U$ for all Mn sites in the structure, instead there are slightly different $U$ values for different Mn sites. This is so because the insertion of Fe in the $\alpha$-MnO$_2$ host lattice leads to structural distortions which break the equivalence of the Mn sites. This approach was also used in Refs.~\cite{Ricca:2019, Ricca:2020} to determine the self-consistent site-dependent Hubbard parameters in perovskites due to the presence of oxygen vacancies. 

\begin{table*}[t]
\renewcommand{\arraystretch}{1.2}
\centering
\resizebox{\columnwidth}{!}{\begin{tabular}{l|c|c|ccccc}
\hline
\multirow{2}{*}{Method}      & \multirow{2}{*}{HP}  & \multirow{2}{*}{Atoms}    & \multicolumn{5}{c}{Magnetic ordering}    \\ \cline{4-8} 
                             &                      &       & A           & B           & C           & D           & E  \\ \hline
\multirow{2}{*}{DFT+$U$}     & \multirow{2}{*}{$U$} & Mn    & $7.01-7.10$ & $6.65-6.72$ & $6.65-6.72$ & $6.54-7.02$ & $6.46-6.48$    \\ 
                             &                      & Fe    & $5.92$      & $6.03$      & $6.02$      & $5.58$      & $5.94$         \\ \hline
\multirow{4}{*}{DFT+$U$+$V$} & \multirow{2}{*}{$U$} & Mn    & $7.24-7.60$ & $6.85-6.95$ & $6.88-6.96$ & $6.78-7.37$ & $6.62-6.73$    \\ 
                             &                      & Fe    & $5.86$      & $6.95$      & $6.94$      & $5.69$      & $6.58$         \\ \cline{2-8} 
                             & \multirow{2}{*}{$V$} & Mn--O & $0.82-1.23$ & $0.75-1.16$ & $0.75-1.18$ & $0.69-1.25$ & $0.64-1.18$    \\ 
                             &                      & Fe--O & $0.53$      & $1.03-1.37$ & $1.05-1.37$ & $0.54-0.63$ & $0.75-1.28$    \\
\hline
\end{tabular}}
\caption{Self-consistent Hubbard parameters (HP) in eV computed using DFPT for five magnetic orderings of the Fe-doped $\alpha$-MnO$_2$: A, B, C, D, and E. The onsite $U$ for Mn($3d$) and Fe($3d$) states and intersite $V$ for Mn($3d$)--O($2p$) and Fe($3d$)--O($2p$) interactions are computed in the frameworks of DFT+$U$ and DFT+$U$+$V$ (PBEsol functional) using L\"owdin-orthogonalized atomic orbitals as Hubbard projector functions.}
\label{tab:Fe-dopedHub_param}
\end{table*}

The Hubbard $U$ parameter for Fe($3d$) states is in the range from roughly 5.6 to 7.0~eV depending on the method used (DFT+$U$ or DFT+$U$+$V$) and on the type of the spin configuration. These values are larger than the value of 5.1~eV which is obtained for the bulk Fe. This shows once again that $U$ is not transferable and it is sensitive to changes in the chemical environment and magnetic interactions. 

The intersite Hubbard $V$ parameters for the Mn--O and Fe--O couples are also listed in Table~\ref{tab:Fe-dopedHub_param}. We can see that $V$ for Mn--O did not change substantially with respect to $V$ values in the pristine $\alpha$-MnO$_2$ case, and they span approximately the same range. Moreover, the $V$ values for Fe--O are similar to those for Mn--O. In order to assess the importance of intersite Hubbard corrections for the Fe-doped $\alpha$-MnO$_2$, in the following we present the ground-state properties computed using DFT+$U$ and DFT+$U$+$V$ with the respective sets of self-consistent site-dependent Hubbard parameters.

\subsubsection{Structural properties}
\label{sec:Fedoped_csp}

Doping of $\alpha$-MnO$_2$ with Fe leads to structural distortions (see Table~S2 in SI). From our calculations we find that in the case of A type doping the lattice preserves the tetragonal symmetry at all levels of theory, while for the B, C, and D types of doping the lattice has a very small monoclinic distortion. In the case of the E type doping, the optimized structure has a triclinic cell at the levels of DFT+$U$ and DFT+$U$+$V$, while at the DFT level the cell is monoclinic. We are aware of only one experimental study of the Fe-doped $\alpha$-MnO$_2$~\cite{song2022fe} showing that the lattice preserves the tetragonal symmetry like in the pristine $\alpha$-MnO$_2$. This seems to suggest that the A type is the most likely spin configuration in the case of the interstitial doping. However, in order to understand better this aspect, further high-resolution experimental investigations of different types of doping of $\alpha$-MnO$_2$ are needed in order to detect possible monoclinic (or triclinic) distortions.

For what concerns the crystal structure parameters, we also observe different trends. In the case of the interstitial doping (A and D), Fe atoms residing in the centers of the $2 \times 2$ channels form bonds with the nearest four O atoms (see Fig.~S2 in SI). Our calculations show that this leads to the reduction of the $a$ and $b$ lattice parameters by $\sim 2\%$ with respect to the pristine $\alpha$-MnO$_2$ case, while the $c$ lattice parameter is increased by $\sim 4\%$ (at the DFT+$U$ and DFT+$U$+$V$ levels of theory). In the case of the substitutional doping (B, C, and E), all lattice parameters are increased by $\sim 1\%$ (at the DFT+$U$ and DFT+$U$+$V$ levels of theory). In contrast, the experimental lattice parameters for the Fe-doped $\alpha$-MnO$_2$ are $a_\mathrm{exp} = b_\mathrm{exp} = 9.83$~\AA\, and $c_\mathrm{exp}=2.86$~\AA~\cite{song2022fe} meaning that there are no noticeable changes with respect to the pristine case (cf. Table~\ref{tab:C2AFMBondLengthAngles}).~\cite{thackeray1997manganese, islam2017carbon, rossouw1992alpha} It should be noted that the uncertainties in the experimental crystal structure parameters are relatively large, and hence more accurate experiments are needed to ascertain possible small changes in the crystal structure due to doping.

We analyzed also the bond lengths and angles for the A and D types of doping (see Fig.~S2 and Table~S3 in SI). We find that due to the Fe doping some Mn--O bonds contract and some expand, and they cover a range of about $1.8-2.1$~\AA. The Fe--O bonds are symmetrical in the A type and they have different lengths in the D type (in the range $2.1-2.2$~\AA). The bond angles Mn--O--Mn are also distorted and their exact values are listed in Table~S3 in SI. Finally, the O--Fe--O angles are $90^\circ$ in the A type and they deviate slightly from $90^\circ$ in the D type, while the Fe--O--Mn angles are in the range $123-126^\circ$. We are not aware of any experimental data for the bond angles and lengths in the Fe-doped $\alpha$-MnO$_2$, hence the accuracy of these predictions cannot be verified at present, which motivates a strong need for new experiments.

\subsubsection{Energetics}
\label{sec:Fedoped_energetics}

As in the case of the pristine $\alpha$-MnO$_2$ (see Sec.~\ref{sec:Energetics}), we compare the total energies of various spin configurations in the case of the Fe-doped $\alpha$-MnO$_2$. It is important to note that since the number of atoms in the simulation cell is not the same for the interstitial and substitutional doping, we can compare the total energies only for spin configurations having the same number of atoms.

As can be seen in Table~\ref{tab:Tot_Energy_diff_interstitial} which is for the interstitial doping, the trends are different at different levels of theory. Namely, DFT predicts the D type to be more energetically favorable than the A type, while DFT+$U$ predicts the opposite trend. Hence, the onsite Hubbard $U$ correction for Mn($3d$) and Fe($3d$) states influences significantly the overall energetics of the system. Remarkably, when the intersite Hubbard $V$ corrections for Mn--O and Fe--O couples are also included, the trend reverses back to the one found in DFT. This shows that intersite Hubbard interactions play a decisive role in stabilizing back the D type in the presence of localized $3d$ electrons on Mn and Fe ions. Hence, the delicate interplay between the onsite localization and intersite hybridization due to covalent bonding is captured by DFT+$U$+$V$ but not by DFT+$U$. Instead, in DFT apparently there is a cancellation of errors which might explain why the D type turns out to be the most energetically favorable as in DFT+$U$+$V$. Therefore, our most accurate level of theory DFT+$U$+$V$ predicts that the D type is more energetically favorable than the A type. This is physically sound since the doping of $\alpha$-MnO$_2$ with Fe is likely to preserve the C2-AFM character of the host; in contrast, switching from C2-AFM to the FM character due to doping would require swapping spins on half of the Mn atoms which has high energetic cost and hence it is less likely.

\begin{table}[t]
\renewcommand{\arraystretch}{1.3}
\centering
\begin{tabular}{c c c c c }
\hline
Magnetic ordering & DFT   & DFT+$U$ & DFT+$U$+$V$ \\ \hline 
A                 & 0.855 & 0       & 0.254 \\
D                 & 0     & 0.072   & 0     \\
\hline
\end{tabular}
\caption{Total energy difference (in eV) in the case of the interstitial doping FeMn$_8$O$_{16}$ (A and D). The zero of energy corresponds to the lowest-energy structure within each level of theory.}
\label{tab:Tot_Energy_diff_interstitial}
\end{table}

In the case of the substitutional doping we find that all levels of theory predict the E type to be the most energetically favorable, as can be seen in Table~\ref{tab:Tot_Energy_diff_substitutional}. Interestingly, for this type of doping the onsite Hubbard $U$ correction alone is able to capture the right energetics of the system at variance with the case of the interstitial doping. Thus, again we find that doping of $\alpha$-MnO$_2$ with Fe preserves the C2-AFM character of the host when making a partial substitution of Mn for Fe.

\begin{table}[t]
\renewcommand{\arraystretch}{1.3}
\centering
\begin{tabular}{c c c c c }
\hline
Magnetic ordering & DFT   & DFT+$U$ & DFT+$U$+$V$ \\ \hline 
B                 & 0.802 & 0.296   & 0.357  \\
C                 & 0.798 & 0.287   & 0.313  \\
E                 & 0     & 0       & 0      \\
\hline
\end{tabular}
\caption{Total energy difference (in eV) in the case of the substitutional doping FeMn$_7$O$_{16}$ (B, C, and E). The zero of energy corresponds to the lowest-energy structure within each level of theory.}
\label{tab:Tot_Energy_diff_substitutional}
\end{table} 

In order to gain more insights about the chemistry of the Fe-doped $\alpha$-MnO$_2$ it would be instructive to compute and compare the defect formation energies for different magnetic orderings shown in Fig.~\ref{fig:Fedoped_structures}.~\cite{Freysoldt:2014, Hoang:2018} This would require carrying out a careful and thorough analysis of calculations using supercells and different charge states of Fe. However, this goes beyond the scope of the current work and will be investigated in the forthcoming study. 

In the following we focus mainly on the D and E type magnetic orderings since these are the lowest-energy configurations within DFT+$U$+$V$.

\subsubsection{Magnetic moment}
\label{sec:magnetic_moments_doped}

\begin{figure}[t]
 \includegraphics[width=0.80\linewidth]{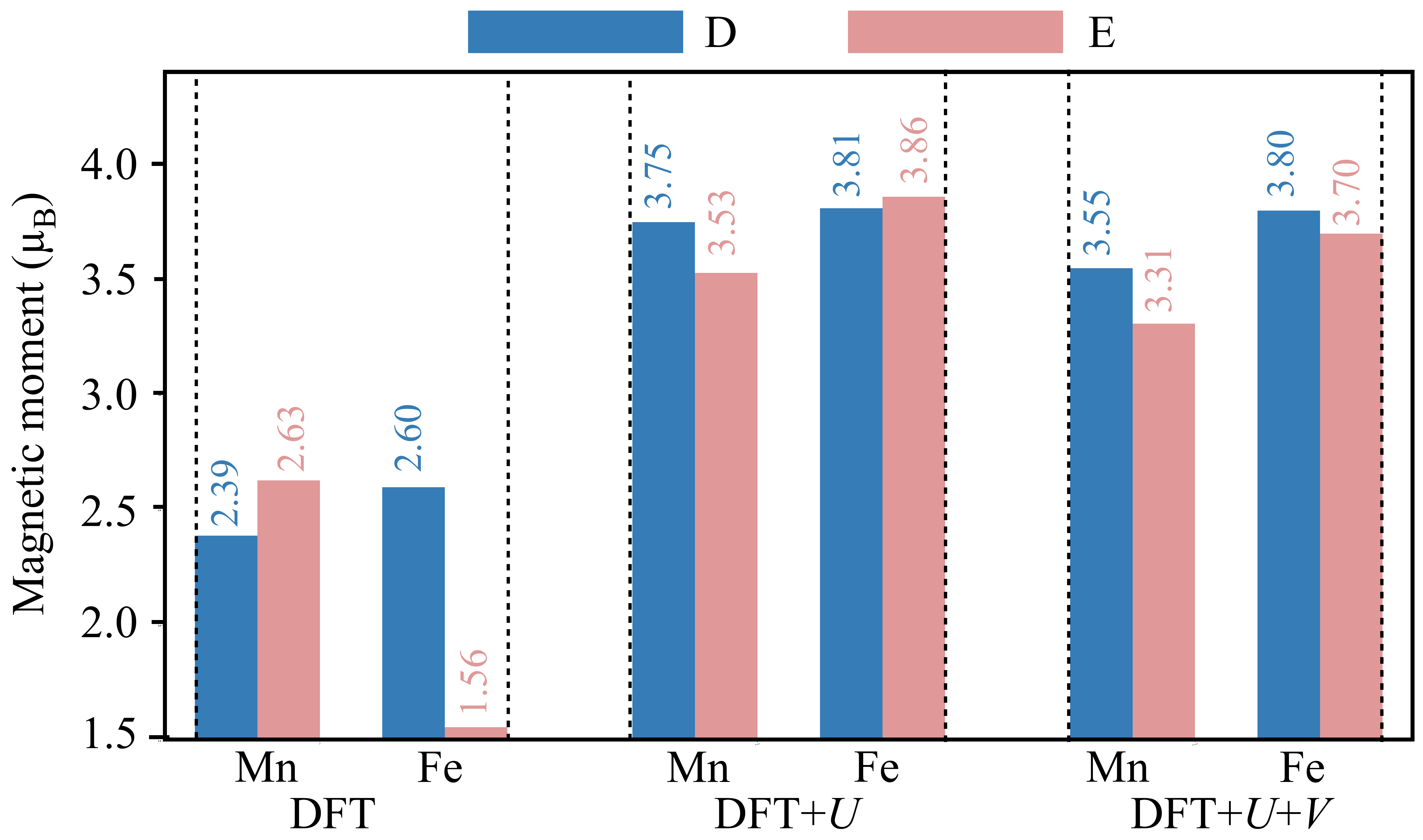}
 \caption{Magnetic moment (in $\mu_\mathrm{B}$) on Mn and Fe atoms in the Fe-doped $\alpha$-MnO$_2$ for two collinear magnetic orderings (D and E types) computed at three levels of theory (DFT, DFT+$U$, and DFT+$U$+$V$) using the PBEsol functional. For each case, the Hubbard parameters $U$ and $V$ were computed using DFPT and are listed in Table~\ref{tab:Fe-dopedHub_param}.}
 \label{fig:Fe-dopedMM}
\end{figure}

The magnetic moments on Mn and Fe atoms for the Fe-doped $\alpha$-MnO$_2$ computed at different levels of theory for the D and E types are shown in Fig.~\ref{fig:Fe-dopedMM} (see Fig.~S3 in SI for the magnetic moments for the A, B, and C types). We are not aware of any experimental reports of magnetic moments for the Fe-doped $\alpha$-MnO$_2$, hence we present the analysis based on a comparison with the magnetic moments in the pristine $\alpha$-MnO$_2$ (see Sec.~\ref{sec:magnetic_moments}).

By comparing Figs.~\ref{fig:Magnetic_moment} and \ref{fig:Fe-dopedMM} we can see that there are no big changes in the magnetic moments on Mn atoms after the Fe doping. Moreover, we still see the same trends as in the case of the pristine $\alpha$-MnO$_2$: DFT and DFT+$U$ give the smallest and the largest magnetic moments, respectively, while DFT+$U$+$V$ gives magnetic moments that are only slightly smaller (by $\sim 0.2 \mu_\mathrm{B}$) than those obtained from DFT+$U$.

The magnetic moments on Fe atoms are very similar within DFT+$U$ and DFT+$U$+$V$ (especially for the D type), in contrast to the magnetic moments on Mn atoms which have larger differences between DFT+$U$ and DFT+$U$+$V$. This is due to the fact that Mn atoms have stronger covalent bonding with ligands,~\cite{cococcioni2019energetics} and hence the magnetic moments are affected stronger by the intersite Hubbard $V$ corrections. Within DFT, the values of magnetic moments on Fe atoms are scattered (for different spin configurations) much more than in the case of DFT+$U$ and DFT+$U$+$V$. This is due to the failure of DFT with standard functionals (LSDA and $\sigma$-GGA) to describe accurately the electronic and magnetic properties of Fe due to large self-interaction errors for $3d$ electrons. Interestingly, the scattering of magnetic moments on Mn atoms in the case of DFT is much smaller than for Fe. Finally, it would be desirable to have the experimental data for the magnetic moments on Mn and Fe in the Fe-doped $\alpha$-MnO$_2$ which would allow us to check the accuracy of computational predictions of this work.

\subsubsection{Projected density of states}
\label{sec:FedopedPDOS}

\begin{figure*}[t]
 \includegraphics[width=0.48\linewidth]{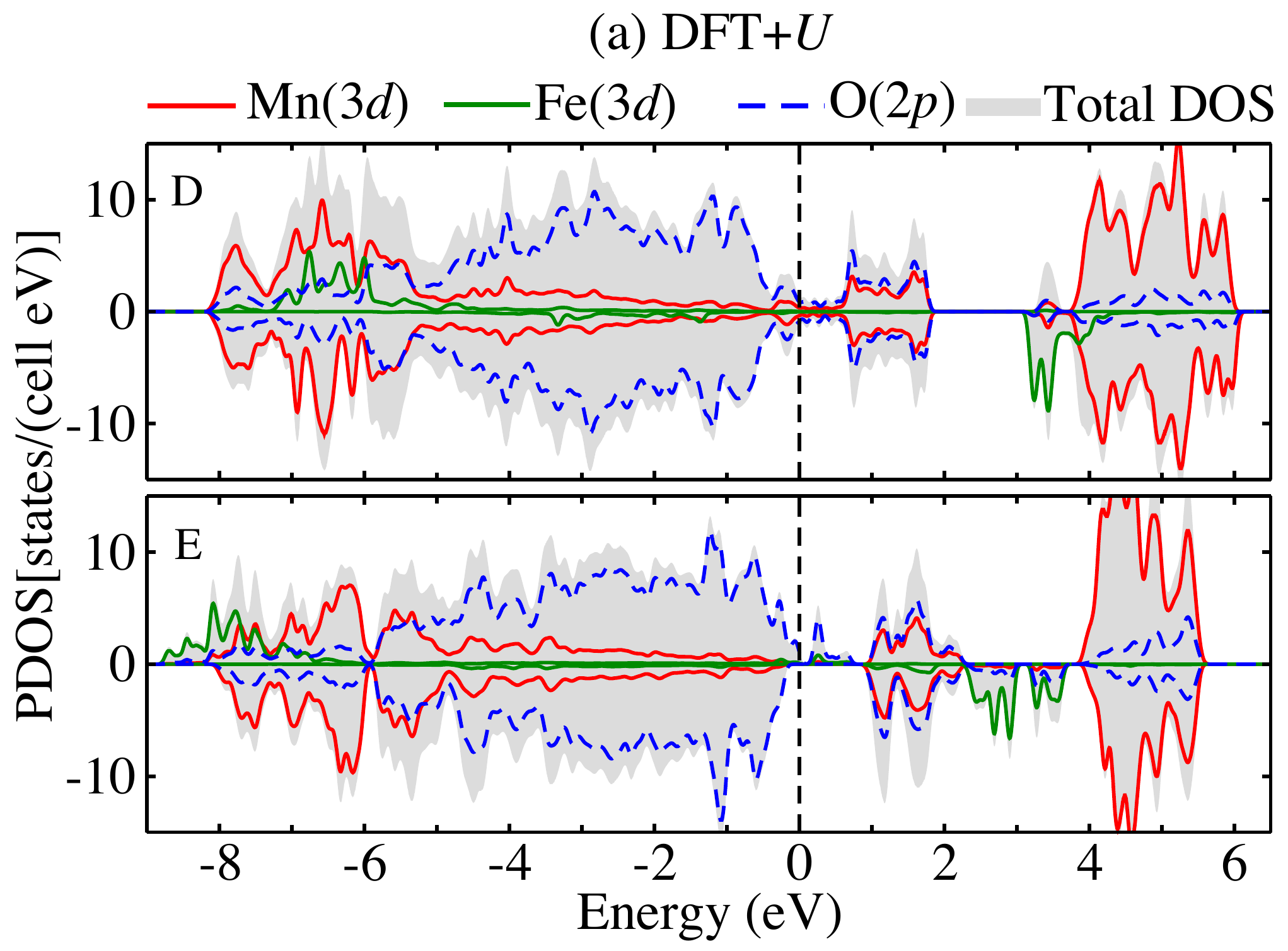}
 \hspace{0.3cm}
 \includegraphics[width=0.48\linewidth]{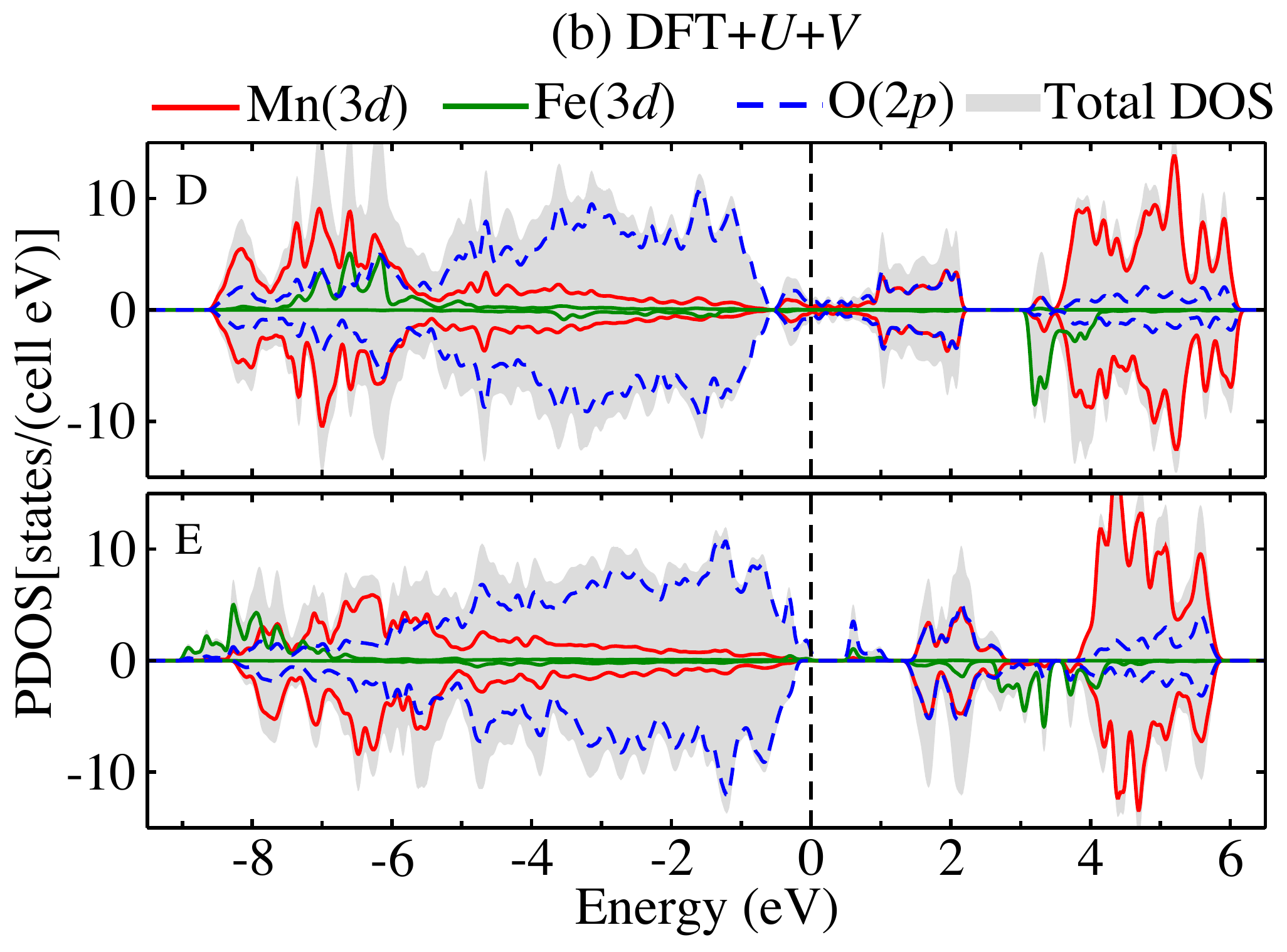}
 \caption{Spin-resolved PDOS and total DOS for two collinear magnetic orderings of the Fe-doped $\alpha$-MnO$_2$ (D and E) computed using the PBEsol functional within (a)~DFT+$U$, and (b)~DFT+$U$+$V$. For each case the Hubbard parameters $U$ and $V$ were computed using DFPT and are listed in Table~\ref{tab:Fe-dopedHub_param}. Upper panels correspond to the spin-up components, while lower panels correspond to the spin-down components. The zero of energy corresponds to the Fermi energy (in the case of metallic ground states) or top of the valence bands (in the case of insulating ground states).}
\label{fig:Fepdos_DFT_Hubbard}
\end{figure*}

After the Fe doping of $\alpha$-MnO$_2$ the electronic structure of this material changes significantly as can be seen in Figs.~\ref{fig:Fepdos_DFT_Hubbard}~(a) and (b) that show the spin-resolved PDOS and total DOS computed using DFT+$U$ and DFT+$U$+$V$ for the D and E types of doping (see Fig.~S4 in SI for the PDOS for the A, B, and C types). We note in passing that the PDOS of the Fe-doped $\alpha$-MnO$_2$ computed using DFT is unreliable [like in the case of the pristine $\alpha$-MnO$_2$, see Fig.~\ref{fig:pdos_pristine}~(a)] due to the overdelocalization of $3d$ electrons of Mn and Fe atoms (see Fig.~S5 in SI). As can be seen in Fig.~\ref{fig:Fepdos_DFT_Hubbard}, the doped material turns out to be a metal in the case of the D type doping, while the E type shows a semiconducting behavior with a small band gap of 0.21~eV in the case of DFT+$U$ and 0.56~eV in the case of DFT+$U$+$V$. From the experimental side, we are aware of Ref.~\cite{Khan:2019} that reports a band gap of 0.30~eV at low Fe concentrations (5 mol\%). It is important to note, though, that a signiﬁcant amount of water was accumulated in the tunnels of the Fe-doped $\alpha$-MnO$_2$ samples in Ref.~\cite{Khan:2019} hence the band gap value reported above might not be very reliable. Nonetheless, both DFT+$U$ and DFT+$U$+$V$ predict a band gap for the E type which is qualitatively consistent with the finding of Ref.~\cite{Khan:2019} In contrast, for the D type we do not find a band gap, though the DOS at the Fermi level is extremely small (see also Fig.~S6 in SI). Therefore, since it is not known which type of doping is dominant in the samples of Ref.~\cite{Khan:2019} (interstitial or substitutional), our DFT+$U$+$V$ calculations for the two best candidates (D and E types) can be considered as satisfactory. Obviously, more experiments on the accurate band gap determination in the Fe-doped $\alpha$-MnO$_2$ are needed.

Of particular interest is to analyze the position of the Fe($3d$) states in the spin-resolved PDOS of the Fe-doped $\alpha$-MnO$_2$. As can be seen in Fig.~\ref{fig:Fepdos_DFT_Hubbard}, in the case of the interstitial doping (D type) these states appear around $-6.5$~eV and $3.5$~eV in the spin-up and spin-down channels, respectively, while in the case of the substitutional doping (E type) they appear around $-8.0$~eV and $2.8$~eV in the spin-up and spin-down channels, respectively. This information could be useful when interpreting the XPS and XANES spectra to determine which type of doping is present in the samples. DFT+$U$ and DFT+$U$+$V$ predict similar position and shape of the peaks due to Fe($3d$) states in the PDOS; the main differences are mainly around the Fermi level (for the A, B, C, and D types) or top of the valence bands (for the E type). Finally, we note that the empty Fe($3d$) states in the spin-down channel have negligible hybridization with the O($2p$) states, while in the spin-up channel the hybridization of the occupied Fe($3d$) states with the O($2p$) states is not very strong. Thus, the intersite $V$ correction for the Fe--O couples is likely not crucial when describing the relative position of the Fe($3d$) and O($2p$) states, while it is relevant for the Mn--O couples which show stronger hybridization especially for the lowest conduction bands.

\subsubsection{Oxidation state}
\label{sec:Oxidation_state}

The OS of Mn and Fe in the Fe-doped $\alpha$-MnO$_2$ is actively discussed in the experimental literature.~\cite{duan2012controlled, hui2012influence, li2017facile, fan2019fe, song2022fe} In Refs.~~\cite{duan2012controlled, hui2012influence, fan2019fe, song2022fe} it is reported that there is a mixture of two types of Mn with the OS of $+4$ and $+3$. It is argued that Mn$^{3+}$ is present due to the appearance of the oxygen vacancies after the Fe doping. The OS of Fe is also reported to be $+4$ and $+3$, with Fe$^{4+}$ being present also due to the oxygen vacancies,~\cite{song2022fe} while in Ref.~\cite{li2017facile} it is argued that the OS of Fe is exclusively $+3$. Computational studies could be useful to verify these experimental findings and assumptions. 

In this work we do not consider oxygen vacancies in the Fe-doped $\alpha$-MnO$_2$ because, first of all, we seek to understand what are the differences in the OS of Mn and Fe for different types of doping (interstitial and substitutional) in the idealized case. We stress that here we consider only neutral Fe dopants. The determination of OS is not a trivial task.~\cite{Raebiger:2008, Resta:2008, Jansen:2008} In particular, using the L\"owdin occupations as a proxy for determining the OS could often be misleading.~\cite{Resta:2008, Timrov:2022b} Here we use the projection-based method of Ref.~\cite{Sit:2011} that has proven to be effective for determining the OS of transition-metal elements in various complex materials.~\cite{Ku:2019, Timrov:2022b} The main idea is based on the count of the eigenvalues of the site-diagonal occupation matrix [see Eq.~(S3) in SI] that are close to unity. Using this approach and our DFT+$U$+$V$ data for the C2-AFM pristine $\alpha$-MnO$_2$ we find that the OS of Mn is $+4$, which agrees with the nominal OS. In Table~S4 in SI we report the eigenvalues of the site-diagonal occupation matrix for the D and E types of the Fe-doped $\alpha$-MnO$_2$. We find that in both cases the OS of Mn is $+4$, while the OS of Fe is $+2$ in the D type and $+4$ in the E type. These findings are physically sound and could be interpreted as follows. In the D type doping, where dopant ions reside in the $2 \times 2$ tunnels, there is a charge transfer from the $4s$ shell of Fe towards four nearest-neighbor O atoms (see Fig.~S2 in SI), while the $3d$ shell of Fe remains intact (thus Fe is in a high-spin state, $d^5(\uparrow)d^1(\downarrow)$). We note though that the $4s$ shell of Fe is not fully empty because the 2 electrons from this shell are actually shared with the nearest four oxygens via hybridizations between the Fe-$4s$ and O-$2p$ orbitals. In contrast, in the E type doping, where the dopant ions partially substitute Mn atoms, Fe tries to adapt itself to the local chemical environment and hence it has the same OS as Mn, i.e. $+4$. As a consequence, in this case Fe ``loses'' 2 electrons from the $4s$ shell and 2 electrons from the $3d$ shell, and thus only 4 electrons remain in the $3d$ shell in the high-spin state, $d^4(\uparrow)d^0(\downarrow)$. The charge transfer occurs from Fe to six O atoms within the FeO$_6$ octahedron. To be more precise, Fe does not lose \textit{fully} these 4 electrons, but it shares them with O atoms by creating covalent bonds, and as a consequence there are fractional atomic occupations of those levels (i.e. the eigenvalues of the occupation matrix have fractional values between 0 and 1).~\cite{Sit:2011} Obviously, the chemistry of the Fe-doped $\alpha$-MnO$_2$ will change in the case when there are oxygen vacancies (see e.g. Ref.~\cite{fan2019fe}) and this will be the topic of future investigations.

\section{Conclusions}
\label{sec:Conclusions}

We have presented a first-principles study of the structural, electronic, and magnetic properties of the pristine and Fe-doped $\alpha$-MnO$_2$ using DFT with extended Hubbard functionals. The onsite and intersite Hubbard parameters were computed self-consistently using DFPT in the basis of L\"owdin-orthogonalized atomic orbitals. We found that DFT+$U$ provides quite accurate description of the ground-state properties of the pristine $\alpha$-MnO$_2$, while the intersite Hubbard $V$ corrections are not decisive though they lead to important quantitative improvements. In particular, we found that the C2-AFM spin configuration is the most energetically favorable one (compared to the FM and other types of the AFM magnetic orderings) in the pristine $\alpha$-MnO$_2$, both within DFT+$U$ and DFT+$U$+$V$. 
This finding shows the superiority of DFT+$U$ and DFT+$U$+$V$ (with $U$ and $V$ determined from linear-response theory) compared to HSE06: the latter fails to predict the antiferromagnetic ground state to be lower in energy than FM and thus contradicts to experiments~\cite{crespo2013electronic}. Overall, the computed crystal structure properties and the band gap of the pristine $\alpha$-MnO$_2$ are in good agreement with experiments, both within DFT+$U$ and DFT+$U$+$V$, while the magnetic moments are somewhat underestimated.

In contrast, in the Fe-doped $\alpha$-MnO$_2$ the onsite Hubbard $U$ corrections alone are unable to predict the correct trends for the interstitial doping and the intersite Hubbard $V$ corrections are crucial. Namely, we find that the interstitial doping preserves the C2-AFM spin configuration of the host lattice only when both onsite $U$ and intersite $V$ Hubbard corrections are included, while for the substitutional doping the onsite Hubbard $U$ correction alone is able to preserve the C2-AFM spin configuration of the host lattice. In addition, the oxidation state of Fe in the two types of doping is found to be different: it is $+2$ in the case of the interstitial doping, and $+4$ in the case of the substitutional doping. The oxidation state of Mn is $+4$ in the pristine and all types of the Fe-doped $\alpha$-MnO$_2$. Finally, we found that the semiconducting character of $\alpha$-MnO$_2$ is preserved in the case of the substitutional doping (E type), while the material becomes metallic with a vanishing DOS at the Fermi level in the case of the interstitial doping (D type).

Finally, this work paves the way for future studies of $\alpha$-MnO$_2$ doped not only with Fe but also other transitional-metal elements or cations in the presence of oxygen vacancies or under strain, as well as the calculation of the defect formation energies.~\cite{Freysoldt:2014, Hoang:2018}

\begin{suppinfo}
Description of the computational method, crystal structure properties, projected density of states, and the population analysis.
\end{suppinfo}

\begin{acknowledgement}
We thank Sokseiha Muy, Nicola Seriani, Ralph Gebauer, and Ankita Mathur for fruitful discussions. R.M. acknowledges support by the IIT Mandi for the HTRA fellowship. I.T. acknowledges support by the NCCR MARVEL, a National Centre of Competence in Research, funded by the Swiss National Science Foundation (Grant number 182892). Computer time was provided by the Holland Computing Centre (University of Nebraska) and by the Swiss National Supercomputing Centre (CSCS) under project No.~s1073.
\end{acknowledgement}


\providecommand{\latin}[1]{#1}
\makeatletter
\providecommand{\doi}
  {\begingroup\let\do\@makeother\dospecials
  \catcode`\{=1 \catcode`\}=2 \doi@aux}
\providecommand{\doi@aux}[1]{\endgroup\texttt{#1}}
\makeatother
\providecommand*\mcitethebibliography{\thebibliography}
\csname @ifundefined\endcsname{endmcitethebibliography}
  {\let\endmcitethebibliography\endthebibliography}{}

\clearpage
\newpage

\begin{center}
{\Large\textbf{Supporting Information for \\ Pivotal Role of Intersite Hubbard Interactions in Fe-Doped $\alpha$-MnO$_2$}}
\end{center}

\renewcommand{\thepage}{S\arabic{page}}  
\setcounter{page}{1}
\renewcommand{\thetable}{S\arabic{table}}  
\setcounter{table}{0}
\renewcommand{\thefigure}{S\arabic{figure}}
\setcounter{figure}{0}
\renewcommand{\thesection}{S\arabic{section}}
\setcounter{section}{0}
\renewcommand{\theequation}{S\arabic{equation}}
\setcounter{equation}{0}

\section{Computational method}

\label{sec:methods}

In this section we briefly discuss the basics of DFT+$U$+$V$~\cite{campo2010extendedSM, himmetoglu2014hubbardSM} and of DFPT for computing Hubbard parameters.~\cite{timrov2018hubbardSM, Timrov:2021SM} All equations can be easily reduced to DFT+$U$ by simply setting $V=0$. 
For the sake of simplicity, the formalism is presented in the framework of norm-conserving (NC) pseudopotentials (PPs) in the collinear spin-polarized case. The generalization to the ultrasoft (US) PPs and the projector augmented wave (PAW) method can be found in ref~\citen{Timrov:2021SM}. Hartree atomic units are used.

\subsection{DFT+$U$+$V$}

In DFT+$U$+$V$, an extended Hubbard correction energy $E_{U+V}$ is added to the approximate DFT energy $E_{\mathrm{DFT}}$:~\cite{campo2010extendedSM}
\begin{equation}
E_{\mathrm{DFT}+U+V} = E_{\mathrm{DFT}} + E_{U+V} .
\label{eq:Edft_plus_u}
\end{equation}
At variance with DFT+$U$ that contains only onsite interactions scaled by $U$, DFT+$U$+$V$ contains also intersite interactions between an atom and its surrounding ligands scaled by $V$. In the case of pristine and Fe-doped $\alpha$-MnO$_2$, the onsite $U$ correction is needed for the Mn($3d$) and Fe($3d$) states, while the intersite $V$ correction is used for Mn($3d$)--O($2p$) and Fe($3d$)--O($2p$) interactions.~\cite{Kulik:2011SM} In the simplified rotationally-invariant formulation~\cite{dudarev1998electronSM} the extended Hubbard correction energy reads:
\begin{equation}
E_{U+V} = \frac{1}{2} \sum_I \sum_{\sigma m m'} 
U^I \left( \delta_{m m'} - n^{II \sigma}_{m m'} \right) n^{II \sigma}_{m' m} - \frac{1}{2} \sum_{I} \sum_{J (J \ne I)}^* \sum_{\sigma m m'} V^{I J} 
n^{I J \sigma}_{m m'} n^{J I \sigma}_{m' m} \,,
\label{eq:Edftu}
\end{equation}
where $I$ and $J$ are the atomic site indices, $m$ and $m'$ are the magnetic quantum numbers associated with a specific angular momentum [$l=2$ for Mn($3d$) and Fe($3d$), and $l=1$ for O($2p$)], $U^I$ and $V^{I J}$ are the effective onsite and intersite Hubbard parameters, and the star in the sum denotes that for each atom $I$ the index $J$ covers all its neighbors up to a given distance. 

The generalized occupation matrices $n^{I J \sigma}_{m m'}$ are based on a projection of the Kohn-Sham (KS) states on localized orbitals $\phi^{I}_{m}(\mathbf{r})$ of neighbor atoms: 
\begin{equation}
n^{I J \sigma}_{m m'} = \sum_{v,\mathbf{k}} f^\sigma_{v,\mathbf{k}}
\braket{\psi^\sigma_{v,\mathbf{k}}}{\phi^{J}_{m'}} \braket{\phi^{I}_{m}}{\psi^\sigma_{v,\mathbf{k}}} \,, 
\label{eq:occ_matrix_0}
\end{equation}
where $v$ and $\sigma$ are the band and spin labels of the KS wavefunctions $\psi^\sigma_{v,\mathbf{k}}(\mathbf{r})$, respectively, $\mathbf{k}$ indicate points in the first Brillouin zone, $f^\sigma_{v,\mathbf{k}}$ are the occupations of the KS states, and $\phi^I_{m}(\mathbf{r}) \equiv \phi^{\gamma(I)}_{m}(\mathbf{r} - \mathbf{R}_I)$ are the localized orbitals centered on the $I$th atom of type $\gamma(I)$ at the position $\mathbf{R}_I$. It is convenient to establish a short-hand notation for the onsite occupation matrix: $n^{I\sigma}_{m m'} \equiv n^{II\sigma}_{m m'}$, which is used in the standard DFT+$U$ approach. The two terms in Eq.~\eqref{eq:Edftu} (i.e., proportional to the onsite $U^{I}$ and intersite $V^{IJ}$ couplings) counteract each other: the onsite term favors localization on atomic sites (thus suppressing hybridization with neighbors), while the intersite term favors hybridized states with components on neighbor atoms. More details about DFT+$U$+$V$ can be found in refs~\citen{campo2010extendedSM, TancogneDejean:2020SM, Lee:2020SM}. Thus, computing $U^I$ and $V^{IJ}$ is crucial to determine the degree of localization of $3d$ electrons on Mn and Fe sites and the degree of hybridization of these $3d$ electrons with $2p$ electrons centered on neighboring O sites. In the next subsection we discuss briefly how these Hubbard parameters can be computed using DFPT.

\subsection{Hubbard parameters from DFPT}
\label{sec:CalcUV_theory}

In Hubbard-corrected DFT the values of Hubbard parameters are not known {\it a~priori}, and hence often these values are chosen empirically such that the final results of simulations match some experimental properties of interest. This is, though, fairly arbitrary and, hence, first-principles calculations of Hubbard parameters are essential and highly desirable. In this paper, we compute $U$ and $V$ from a generalized piece-wise linearity condition imposed through linear-response theory,~\cite{cococcioni2005linearSM} based on density-functional perturbation theory (DFPT).~\cite{timrov2018hubbardSM, Timrov:2021SM} DFPT has proven to be effective for determining Hubbard parameters for a variety of systems with complex magnetic properties.~\cite{Ricca:2019SM, floris2020hubbardSM, Sun:2020SM, Zhou:2021SM} Within this framework, the Hubbard parameters are defined as:~\cite{cococcioni2005linearSM}
\begin{equation}
U^I = \left(\chi_0^{-1} - \chi^{-1}\right)_{II} \,, \quad 
V^{IJ} = \left(\chi_0^{-1} - \chi^{-1}\right)_{IJ} \,,
\label{eq:Ucalc}
\end{equation}
where $\chi_0$ and $\chi$ are the bare and self-consistent susceptibilities which measure the response of atomic occupations to shifts in the potential acting on individual Hubbard manifolds. $\chi$ is defined as $\chi_{IJ} = \sum_{m\sigma} \left(dn^{I \sigma}_{mm} / d\alpha^J\right)$, where $\alpha^J$ is the strength of the perturbation of electronic occupations of the $J$th site, and it is computed at self-consistency of the DFPT calculation, while $\chi_0$ has a similar definition but it is computed before the self-consistent re-adjustment of the Hartree and exchange-correlation potentials.~\cite{timrov2018hubbardSM} The response of the occupation matrix is computed in a unit cell as:
\begin{equation}
\frac{dn^{I \sigma}_{mm'}}{d\alpha^J} = \frac{1}{N_{\mathbf{q}}}\sum_{\mathbf{q}}^{N_{\mathbf{q}}} e^{i\mathbf{q}\cdot(\mathbf{R}_{l} - \mathbf{R}_{l'})}\Delta_{\mathbf{q}}^{s'} n^{s \sigma}_{mm'} \,,
\label{eq:dnq}
\end{equation}
where $\mathbf{q}$ is the wavevector of the monochromatic perturbation, $N_\mathbf{q}$ is the total number of $\mathbf{q}$'s, $\Delta_{\mathbf{q}}^{s'} n^{s \sigma}_{mm'}$ is the lattice-periodic response of atomic occupations to a $\mathbf{q}$-specific monochromatic perturbation, $I\equiv(l,s)$ and $J\equiv(l',s')$, where $s$ and $s'$ are the atomic indices in unit cells while $l$ and $l'$ are the unit cell indices, $\mathbf{R}_l$ and $\mathbf{R}_{l'}$ are the Bravais lattice vectors. The $\mathbf{q}$ grid must be chosen dense enough to make the atomic perturbations decoupled from their periodic replicas. More details about this approach can be found in refs~\citen{timrov2018hubbardSM, Timrov:2021SM}. It is important to note that the main advantage of DFPT is that it does not require the usage of computationally expensive supercells contrary to the original linear-response formulation of ref~\citen{cococcioni2005linearSM}. 

Finally, it is very important to remind that the values of the computed Hubbard parameters are strongly dependent on the type of Hubbard projector functions $\phi^I_{m}(\mathbf{r})$ that are used in Eq.~\eqref{eq:occ_matrix_0}. In this work we use the atomic orbitals orthogonalized using the L\"owdin method.~\cite{lowdin1950nonSM, mayer2002lowdinSM} In fact, it has been shown in previous studies that Hubbard-corrected DFT with this type of Hubbard projectors and respective Hubbard parameters computed using DFPT gives an accurate description of various materials' properties.~\cite{Ricca:2020SM, Timrov:2020cSM, kirchner2021extensiveSM, Xiong:2021SM, Timrov:2022b} In particular, in ref~\citen{mahajan2021importanceSM} we have shown that the most accurate description of the structural, electronic, and magnetic properties of $\beta$-MnO$_2$ is obtained using L\"owdin-orthogonalized atomic orbitals as Hubbard projectors.


\newpage

\section{Crystal structure properties and projected density of states of the pristine $\alpha$-MnO$_2$}
\label{sec:pristine-csp}

\begin{table*}[h!]
\begin{center}
\renewcommand{\arraystretch}{1.1}
\resizebox{\columnwidth}{!}{\begin{tabular}{ccccc}
 \hline
 \parbox{3.2cm}{\centering Magnetic ordering} & 
 \parbox{3.2cm}{\centering CSP} & 
 \parbox{3.2cm}{\centering DFT}	& 
 \parbox{3.2cm}{\centering DFT+$U$}	& 
 \parbox{3.2cm}{\centering DFT+$U$+$V$} \\
 \hline 
\multirow{4}{*}{FM} & $a$ (\AA) & 9.71	& 9.93	& 9.84 \\ 
 & $c$ (\AA) & 2.94 & 2.95 & 	2.92 \\ 
 & $V$ (\AA$^3$) & 267.47	& 291.14 & 282.34 \\
 & $\Delta V$ ($\%$) &  -2.56 & 6.06 & 2.86 \\
 \hline
\multirow{4}{*}{A2-AFM} & $a$ (\AA) & 9.71 & 9.86	& 9.81 \\ 
 & $c$ (\AA) &2.84	& 2.94 & 2.91 \\ 
 & $V$ (\AA$^3$) & 267.5	& 285.42	& 279.83 \\
 & $\Delta V$ ($\%$) & -2.55	& 3.98	& 1.94 \\
 \hline
\multirow{4}{*}{C-AFM} & $a$ (\AA) & 9.66	& 9.81	& 9.76  \\ 
 & $c$ (\AA) & 2.83 & 2.93 & 2.90 \\ 
 & $V$ (\AA$^3$) & 264.44 &	281.55 & 276.76 \\
 & $\Delta V$ ($\%$) & -3.66 &	2.57 &	0.82	 \\
 \hline
\multirow{4}{*}{C2-AFM} & $a$ (\AA) & 9.66	& 9.85	& 9.78	 \\ 
 & $c$ (\AA) & 2.83 & 	2.93	& 2.91 \\
  & $V$ (\AA$^3$) & 263.91 & 284.61 &	278.26	 \\
& $\Delta V$ ($\%$) & -3.86	& 3.68	& 1.37 \\
 \hline
\multirow{4}{*}{G-AFM} & $a$ (\AA) & 9.69	& 9.83 &	9.78\\ 
 & $c$ (\AA) & 2.82	& 2.92 &	2.90 \\ 
  & $V$ (\AA$^3$) & 265.16	& 282.57 &	277.41 \\
& $\Delta V$ ($\%$) & -3.40	& 2.94	& 1.06 \\
 \hline
\end{tabular}}
\caption{Crystal structure parameters (CSP) of the pristine $\alpha$-MnO$_2$ (see Fig.~1 in the main text): lattice parameters $a$ and $c$ (in \AA), volume $V$ (in \AA$^3$) corresponding to the 24-atoms unit cell and its deviation $\Delta V$ (in $\%$) from the experimental value 274.5~\AA$^3$ of ref~\citen{rossouw1992alphaSM}. All theoretical predictions preserve the tetragonal symmetry (hence $a=b$ and we report only $a$). The results are presented for five collinear magnetic orderings (FM, A2-AFM, C-AFM, C2-AFM, and G-AFM) computed at three level of theory (see Fig.~2 in the main text): DFT, DFT+$U$, and DFT+$U$+$V$ (PBEsol functional). Hubbard-corrected results are obtained using L\"owdin-orthogonalized atomic orbitals as Hubbard projector functions and the respective Hubbard parameters (see Table~1 in the main text). The experimental CSP for the pristine $\alpha$-MnO$_2$ (tetragonal structure) are $a_\mathrm{exp}=9.75$~\AA, $c_\mathrm{exp}=2.86$~\AA, and $V_\mathrm{exp}=272.0$~\AA$^3$ according to ref~\citen{thackeray1997manganeseSM}; $a_\mathrm{exp}=9.84$~\AA, $c_\mathrm{exp}=2.86$~\AA, and $V_\mathrm{exp}=276.9$~\AA$^3$ according to ref~\citen{islam2017carbonSM}; $a_\mathrm{exp}=9.79$~\AA, $c_\mathrm{exp}=2.86 \, (2.87)$~\AA, and $V_\mathrm{exp}=274.1 \, (274.5)$~\AA$^3$ according to ref~\citen{rossouw1992alphaSM}.}

\label{tab:csp_pure}
\end{center}
\end{table*}

\begin{figure}[h!]
\begin{center}
   \includegraphics[width=0.47\linewidth]{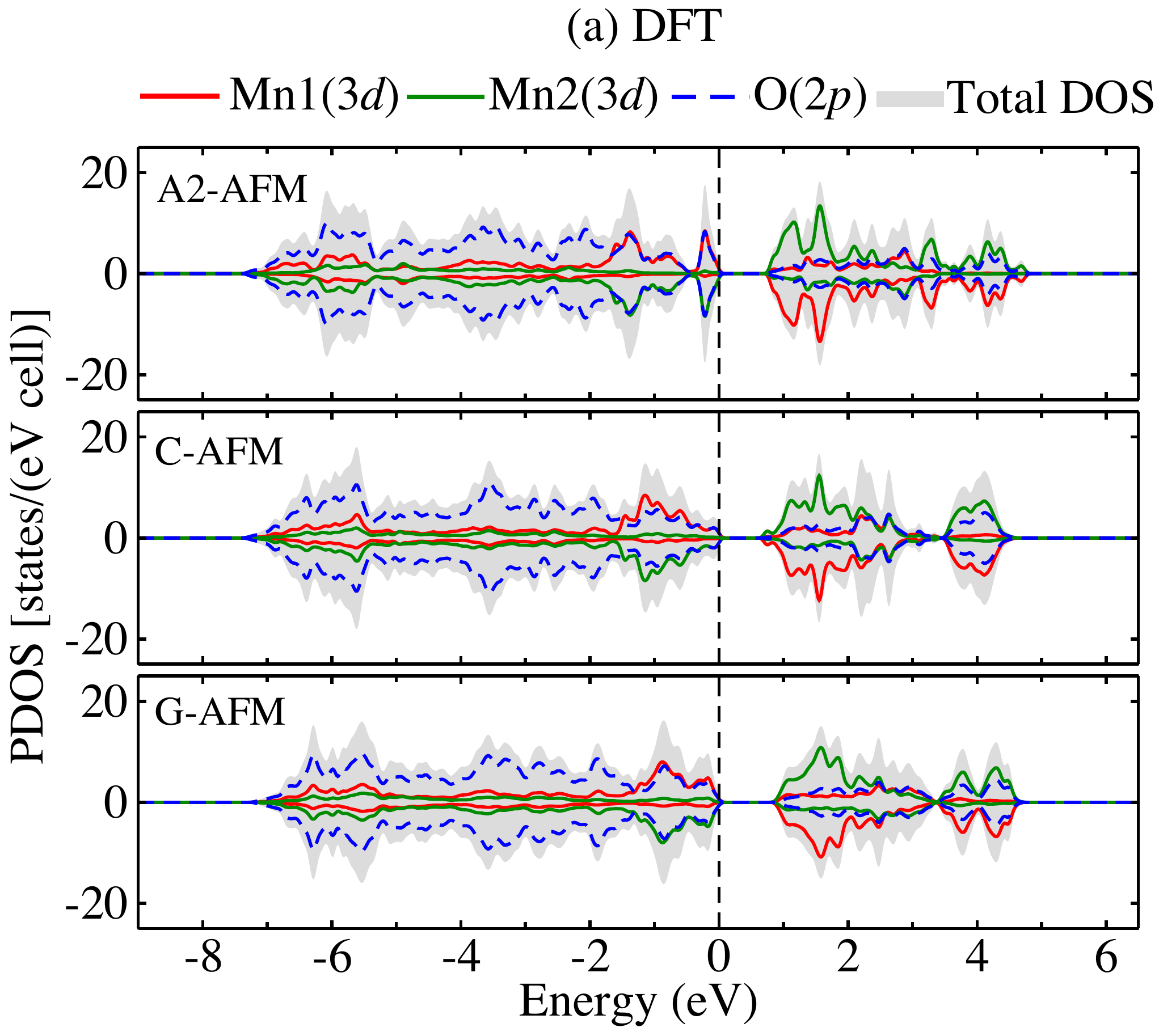} \hspace{0.4cm}
   \includegraphics[width=0.47\linewidth]{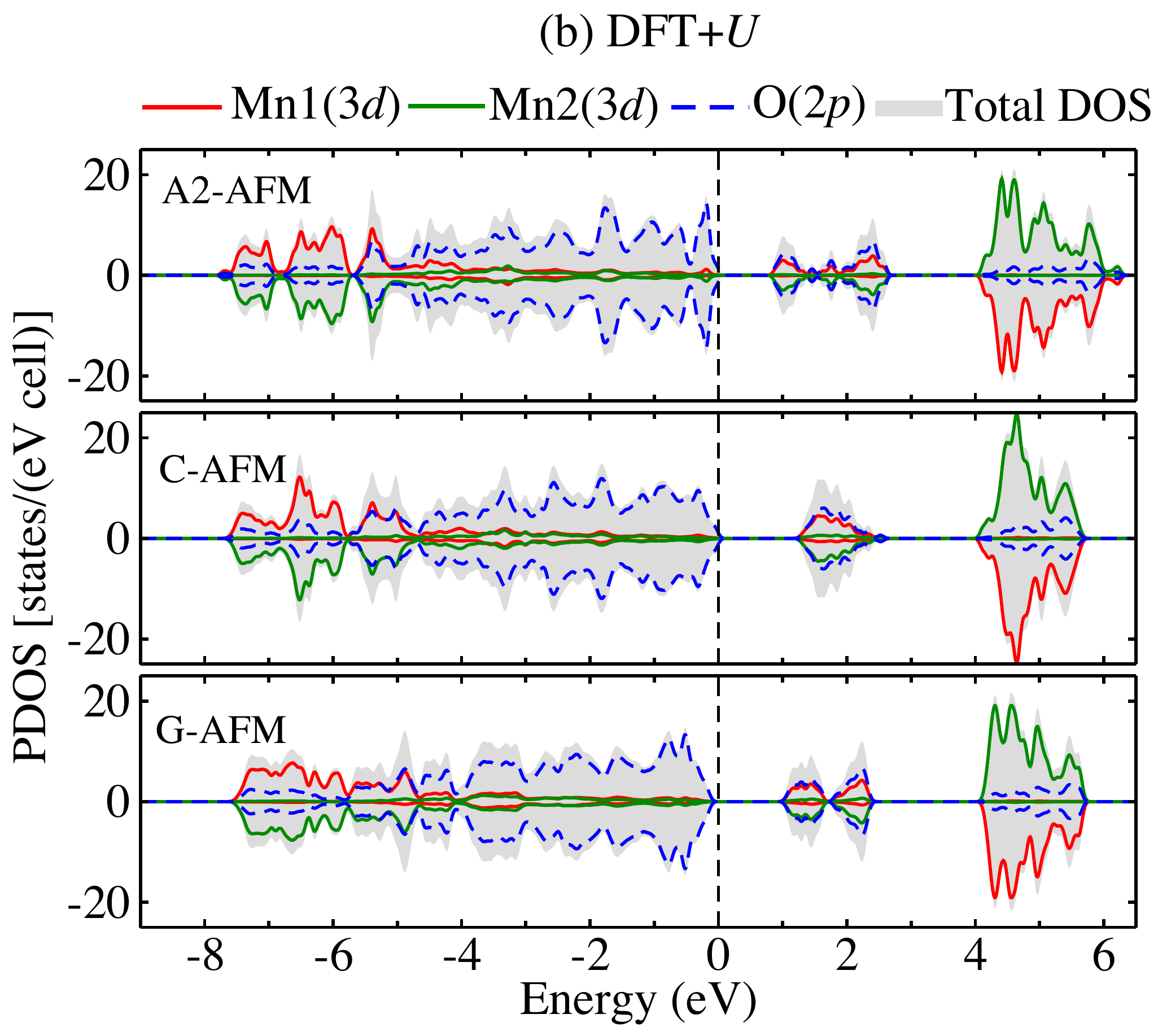}
   \includegraphics[width=0.47\linewidth]{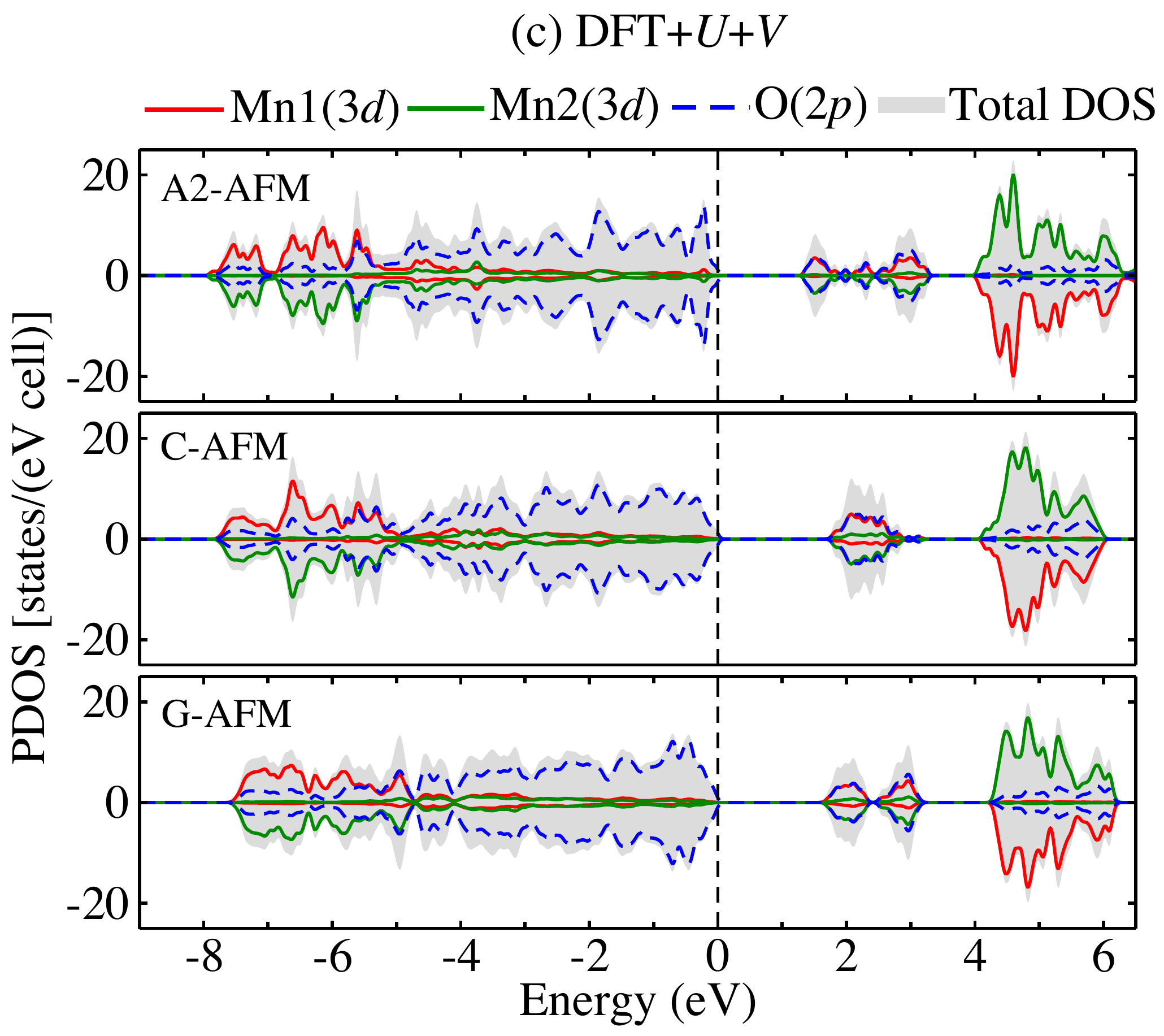}
   \caption{Spin-resolved PDOS and total DOS for the A2-AFM, C-AFM, and G-AFM magnetic orderings of the pristine $\alpha$-MnO$_2$ computed at three levels of theory using the PBEsol functional: (a)~DFT, (b)~DFT+$U$, and (c)~DFT+$U$+$V$. For each case, the Hubbard parameters $U$ and $V$ were computed using DFPT using L\"owdin-orthogonalized atomic orbitals and are listed in Table~1 of the main text. The zero of energy corresponds to the top of the valence bands. The intensity of PDOS and total DOS was rescaled to the 24-atoms unit cell for G-AFM.}
   \label{fig:Fepdos_DFT}
\end{center}
\end{figure}


\clearpage
\newpage

\section{Crystal structure properties, projected density of states, population analysis for the Fe-doped $\alpha$-MnO$_2$}

\begin{table*}[h!]
\begin{center}
\centering
\renewcommand{\arraystretch}{0.8}
\resizebox{\columnwidth}{!}{\begin{tabular}{ccccc}
 \hline
 \parbox{3.2cm}{\centering Magnetic ordering} &  
 \parbox{3.2cm}{\centering CSP}               &  
 \parbox{3.2cm}{\centering DFT}               & 
 \parbox{3.2cm}{\centering DFT+$U$}           &
 \parbox{3.2cm}{\centering DFT+$U$+$V$}       \\ 
 \hline 
\multirow{7}{*}{A} & $a$  & 9.49	& 9.68	& 9.62  \\ 
  & $b$  & 9.49 & 9.68 & 	9.62 \\ 
  & $c$ & 2.84	& 3.07	& 3.03	  \\
  & $\alpha$ & 90.00	&  90.00	& 90.00	  \\
  & $\beta$ & 90.00	& 90.00	& 90.00	  \\
  & $\gamma$ & 90.00	& 90.00	& 90.00	  \\
  & $V$ & 255.80	& 287.29	& 280.38	  \\
 \hline
\multirow{7}{*}{B} & $a$ & 9.70  &  9.94	& 9.86 \\ 
  & $b$  &9.69	& 9.95 & 9.87 \\ 
  & $c$  &  2.83	&  2.96	&  2.93    \\
  & $\alpha$ & 90.00 & 90.00	& 90.00	  \\
  & $\beta$ & 90.00	& 90.00	& 90.00	  \\
  & $\gamma$ & 89.81 & 89.79	& 89.87	  \\
  & $V$ & 266.30	& 292.73	& 284.81	  \\
 \hline
\multirow{7}{*}{C} & $a$  & 9.68	& 9.96	& 9.87  \\ 
  & $b$  & 9.70 & 9.94 & 9.86  \\ 
  & $c$  &  2.83	 & 2.96	  & 2.93   \\
  & $\alpha$ & 90.00 & 90.00	& 90.00	  \\
  & $\beta$ & 90.00 	& 90.00	& 90.00	  \\
  & $\gamma$ & 90.23	& 90.17	& 90.09	  \\
  & $V$ & 266.08	& 292.99	& 284.79	  \\
 \hline
\multirow{7}{*}{D} & $a$  & 9.49	& 9.60 &9.56	 \\  
  & $b$  & 9.48	& 9.60 &	9.56	 \\ 
  & $c$  & 2.82	& 3.06 &	3.03	  \\
  & $\alpha$ & 90.00	& 90.00	& 90.00	  \\
  & $\beta$ & 90.00	& 90.00	& 90.00	  \\
  & $\gamma$ & 89.75	& 89.54	& 89.87	  \\
  & $V$ & 253.80	& 281.92	& 276.41	  \\
 \hline
\multirow{7}{*}{E} & $a$  & 9.63	& 9.89	& 9.84 	 \\  
  & $b$  & 9.64 & 	9.87	&  9.77 \\
  & $c$ & 2.83	& 2.94 &	2.91	  \\
  & $\alpha$ & 90.00	& 89.86	& 89.79	  \\
  & $\beta$ & 90.00	& 89.50	& 89.40	  \\
  & $\gamma$ & 90.21	& 90.09	& 89.88	  \\
  & $V$ & 262.36	& 286.89	& 280.14	  \\
 \hline
\end{tabular}}
\caption{Crystal structure parameters (CSP) of the Fe-doped $\alpha$-MnO$_2$: lattice parameters $a$, $b$, and $c$ (in \AA), angles $\alpha$, $\beta$, $\gamma$ (in degrees), and the volume $V$ (in \AA$^3$). The results are presented for five collinear magnetic orderings (A, B, C, D, and E - see Fig.~7 in the main text) computed at three level of theory: DFT, DFT+$U$, and DFT+$U$+$V$ (PBEsol functional). Hubbard-corrected results are obtained using L\"owdin-orthogonalized atomic orbitals as Hubbard projector functions and the respective Hubbard parameters (see Table~3 in the main text). The experimental CSP for the Fe-doped $\alpha$-MnO$_2$ are $a_\mathrm{exp}=9.83$~\AA, $b_\mathrm{exp}=9.83$~\AA, $c_\mathrm{exp}=2.86$~\AA, and $V_\mathrm{exp}=276.4$.~\cite{song2022feSM}}
\label{tab:csp_Fedoped}
\end{center}
\end{table*}

\begin{table*}[h!]
\renewcommand{\arraystretch}{1.2}
\centering
\resizebox{\columnwidth}{!}{\begin{tabular}{c|ccc|ccc}
\hline
  \multirow{2}{*}{\parbox{1.0cm}{\centering CSP}} & \multicolumn{3}{c}{\centering A} & \multicolumn{3}{c}{\centering D} \\ \cline{2-7}
                       & \parbox{2.0cm}{\centering DFT} & \parbox{2.0cm}{\centering DFT+$U$} & \parbox{2.0cm}{\centering DFT+$U$+$V$} & \parbox{2.5cm}{\centering DFT} & \parbox{2.5cm}{\centering DFT+$U$} & \parbox{2.5cm}{\centering DFT+$U$+$V$} \\
\hline 
  Mn--O     & $1.83-2.00$ & $1.86-2.10$ & $1.85-2.09$ & $1.81-2.02$     & $1.86-2.13$     & $1.85-2.11$      \\
  Fe--O     & $2.18$      & $2.16$      & $2.17$      & $2.16-2.19$     & $2.13-2.21$     & $2.13-2.21$      \\
$\theta_1$  & $130.02$    & $127.08$    & $127.31$    & $130.40-130.41$ & $126.69$        & $127.07$         \\
$\theta_2$  & $97.87$     & $98.90$     & $98.77$     & $98.34-98.63$   & $98.42-99.04$   & $98.51-99.08$    \\
$\theta_3$  & $90.73$     & $89.92$     & $90.32$     & $90.57-90.78$   & $90.48-91.14$   & $90.50-91.13$    \\
$\phi_1$    & $90.00$     & $90.00$     & $90.00$     & $90.07$         & $90.14$         & $90.05$          \\
$\phi_2$    & $125.07$    & $125.20$    & $125.16$    & $124.12-125.12$ & $125.98-126.43$ & $125.73-126.32$  \\
$\phi_3$    & $124.07$    & $124.00$    & $123.93$    & $124.23-124.61$ & $122.91-123.62$ & $123.18-123.73$  \\ 
$\phi_4$    & $90.00$     & $90.00$     & $90.00$     & $89.93$         & $89.86-89.87$   & $89.95$          \\
\hline
\end{tabular}}
\caption{Crystal structure parameters (CSP) for the A and D types of the Fe-doped $\alpha$-MnO$_2$ at the levels of DFT, DFT+$U$, and DFT+$U$+$V$. Bond lengths are shown for Mn--O and Fe--O couples, and the angles $\theta_1$--$\theta_3$ and $\phi_1$--$\phi_4$ are shown in Fig.~\ref{fig:A_and_D_Fe-doped}.}
\label{tab:A_and_D_Fe-doped}
\end{table*}

\begin{figure}[h!]
\begin{center}
   \includegraphics[width=0.45\linewidth]{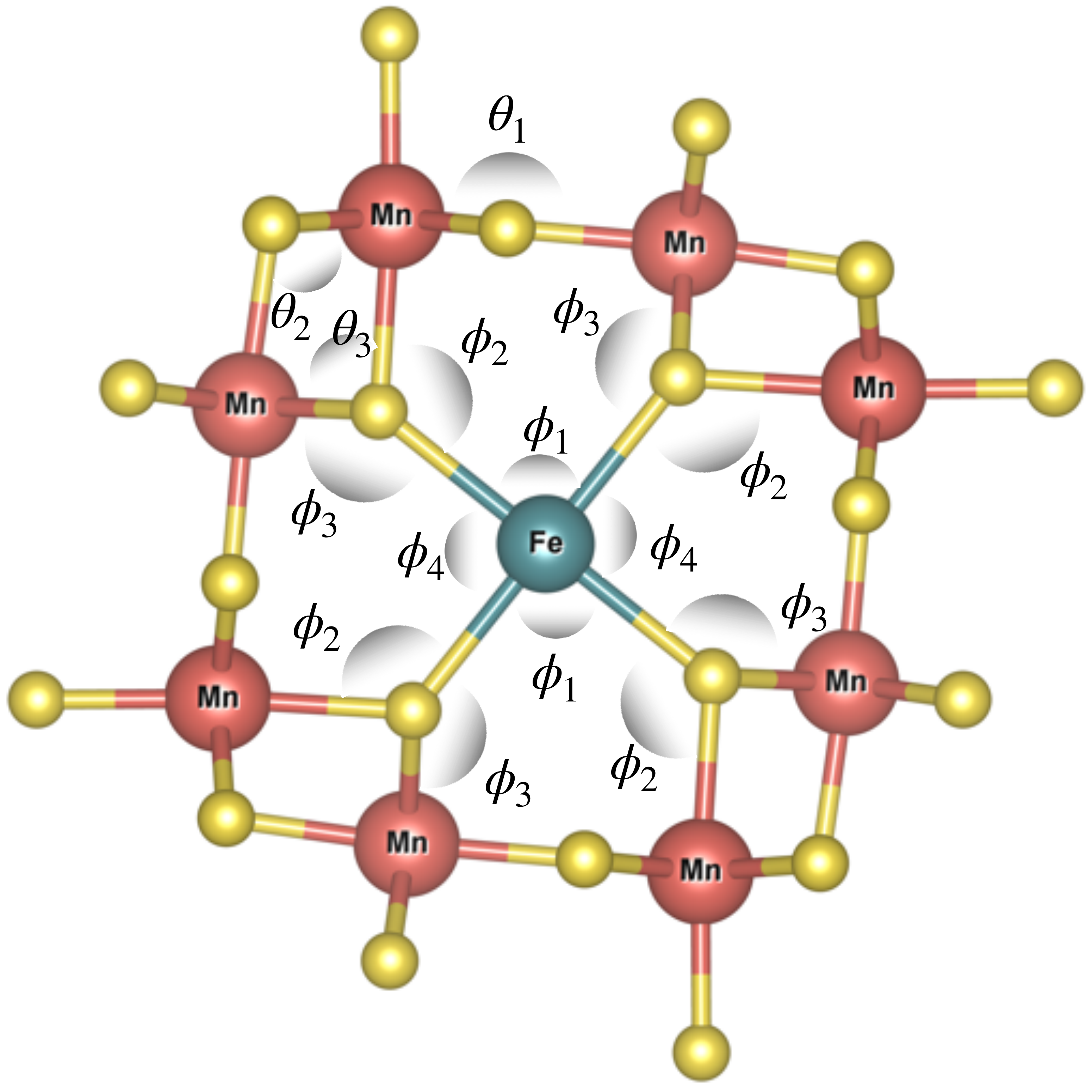}
   \hspace{0.4cm}
   \includegraphics[width=0.45\linewidth]{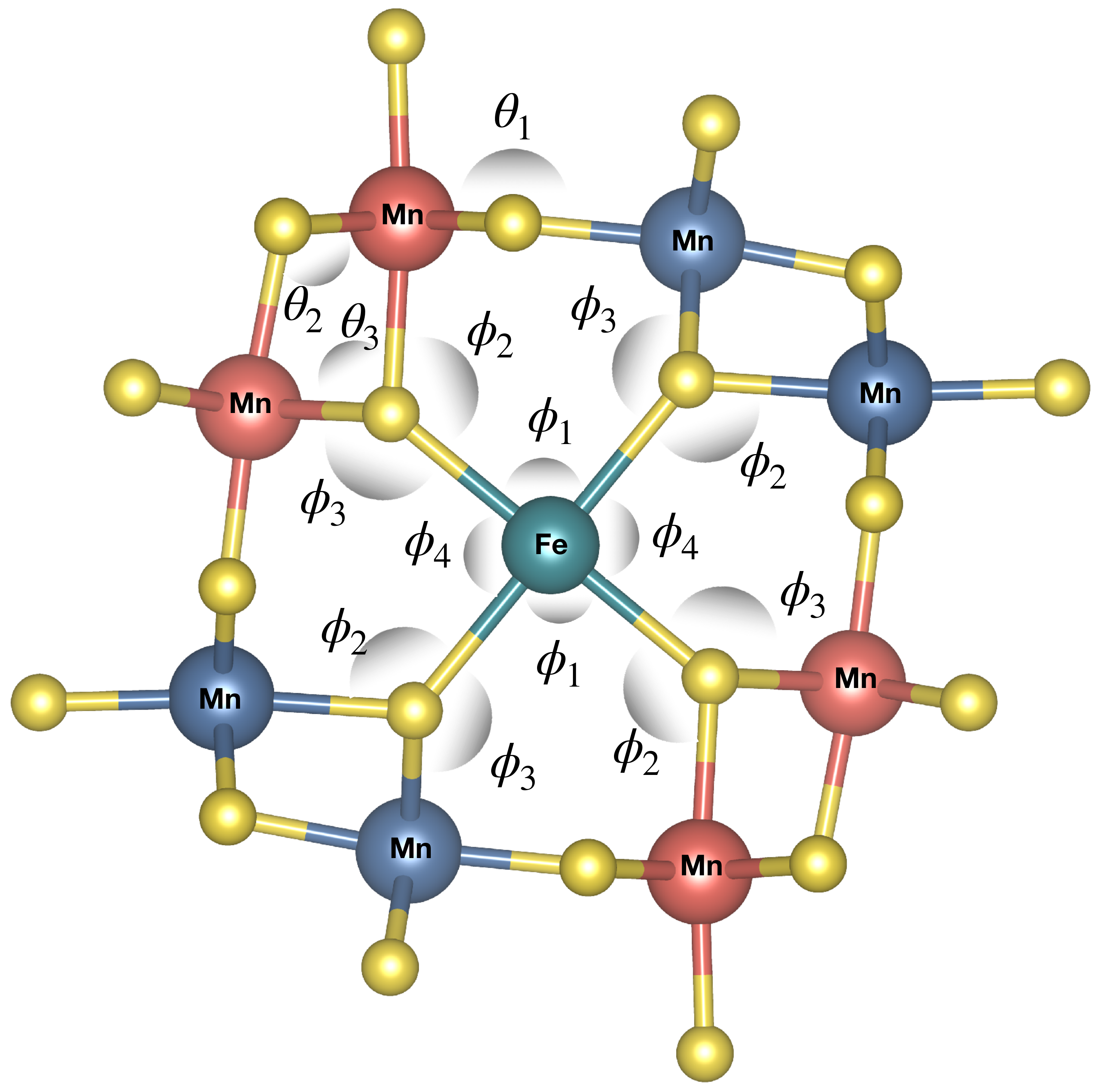}
   \caption{Bond angles in the Fe-doped $\alpha$-MnO$_2$ for the A type (left panel) and D type (right panel) spin configurations (see Fig.~7 in the main text). Light brown and blue colors correspond to Mn atoms with spin-up and spin-down alignments, respectively, while light green color corresponds to the Fe atom with the spin-up alignment. The oxygen atoms are shown in yellow color.}
   \label{fig:A_and_D_Fe-doped}
\end{center}
\end{figure}

\begin{figure}[h!]
\begin{center}
   \includegraphics[width=0.80\linewidth]{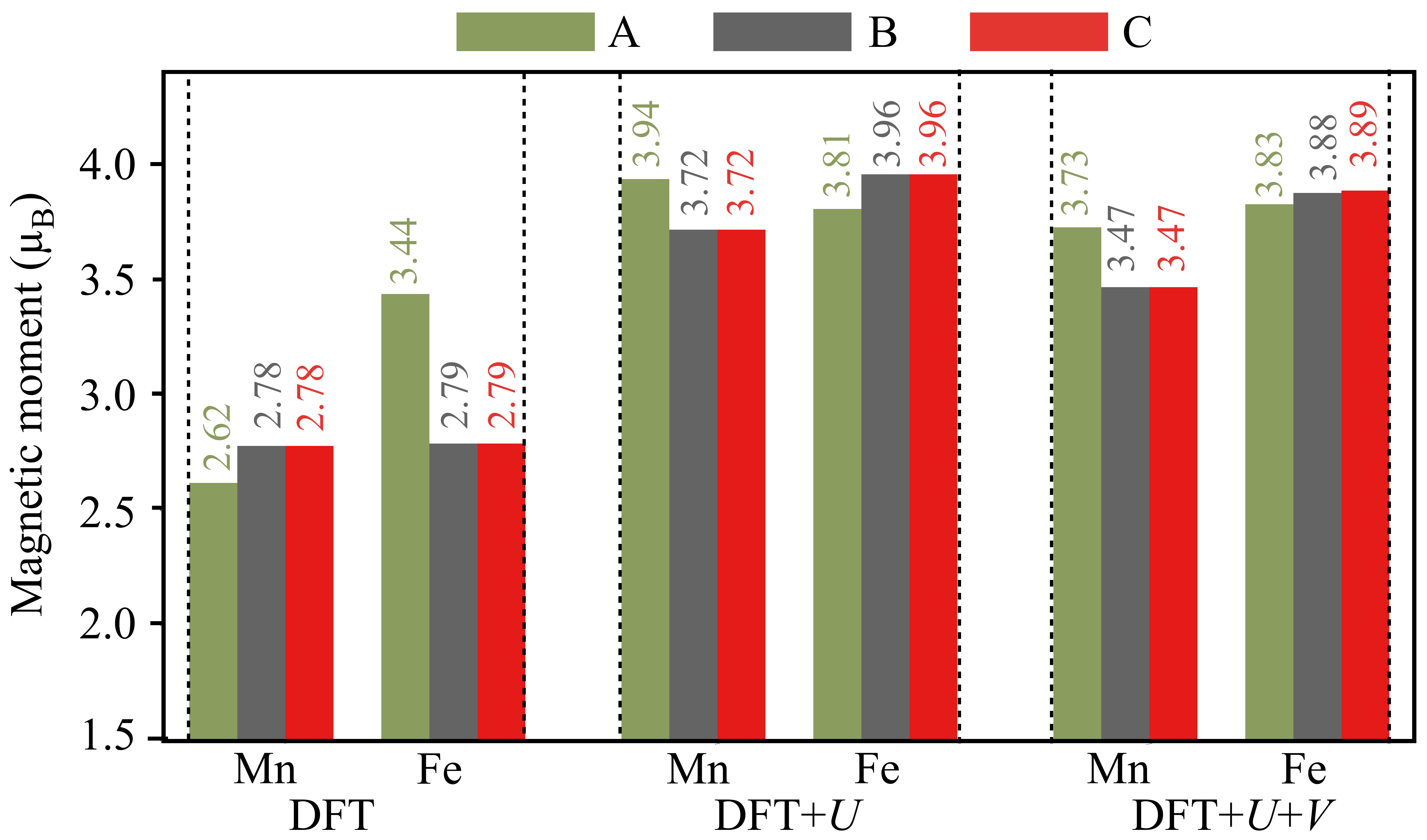}
   \caption{Magnetic moment (in $\mu_\mathrm{B}$) on Mn and Fe atoms in the Fe-doped $\alpha$-MnO$_2$ for three collinear magnetic orderings (A, B, and C types) computed at three levels of theory (DFT, DFT+$U$, and DFT+$U$+$V$) using the PBEsol functional. For each case, the Hubbard parameters $U$ and $V$ were computed using DFPT and are listed in Table~3 in the main text.}
\label{fig:Fepdos_DFT}
\end{center}
\end{figure}

\begin{figure}[h!]
\begin{center}
   \includegraphics[width=0.47\linewidth]{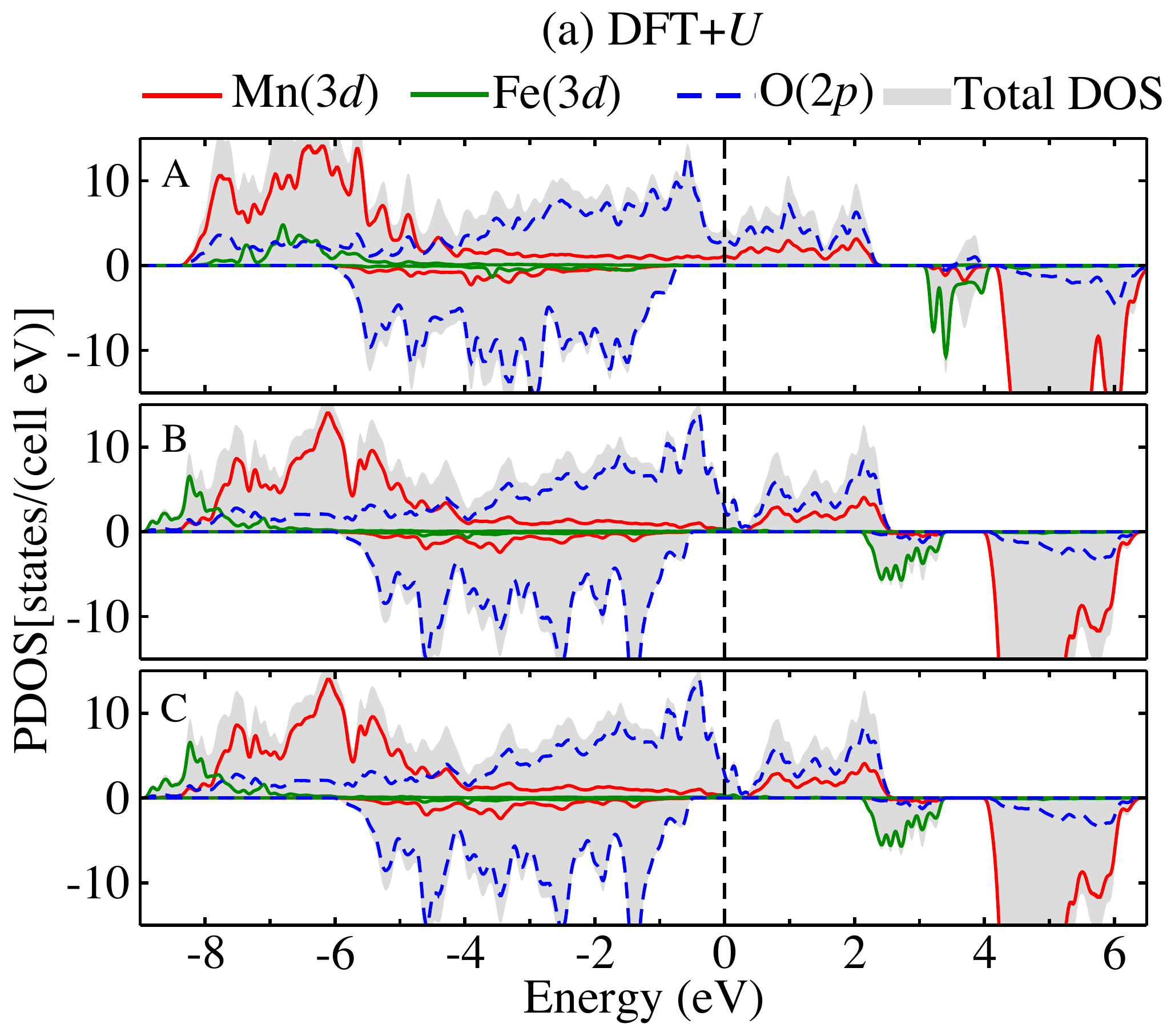} \hspace{0.4cm}
   \includegraphics[width=0.47\linewidth]{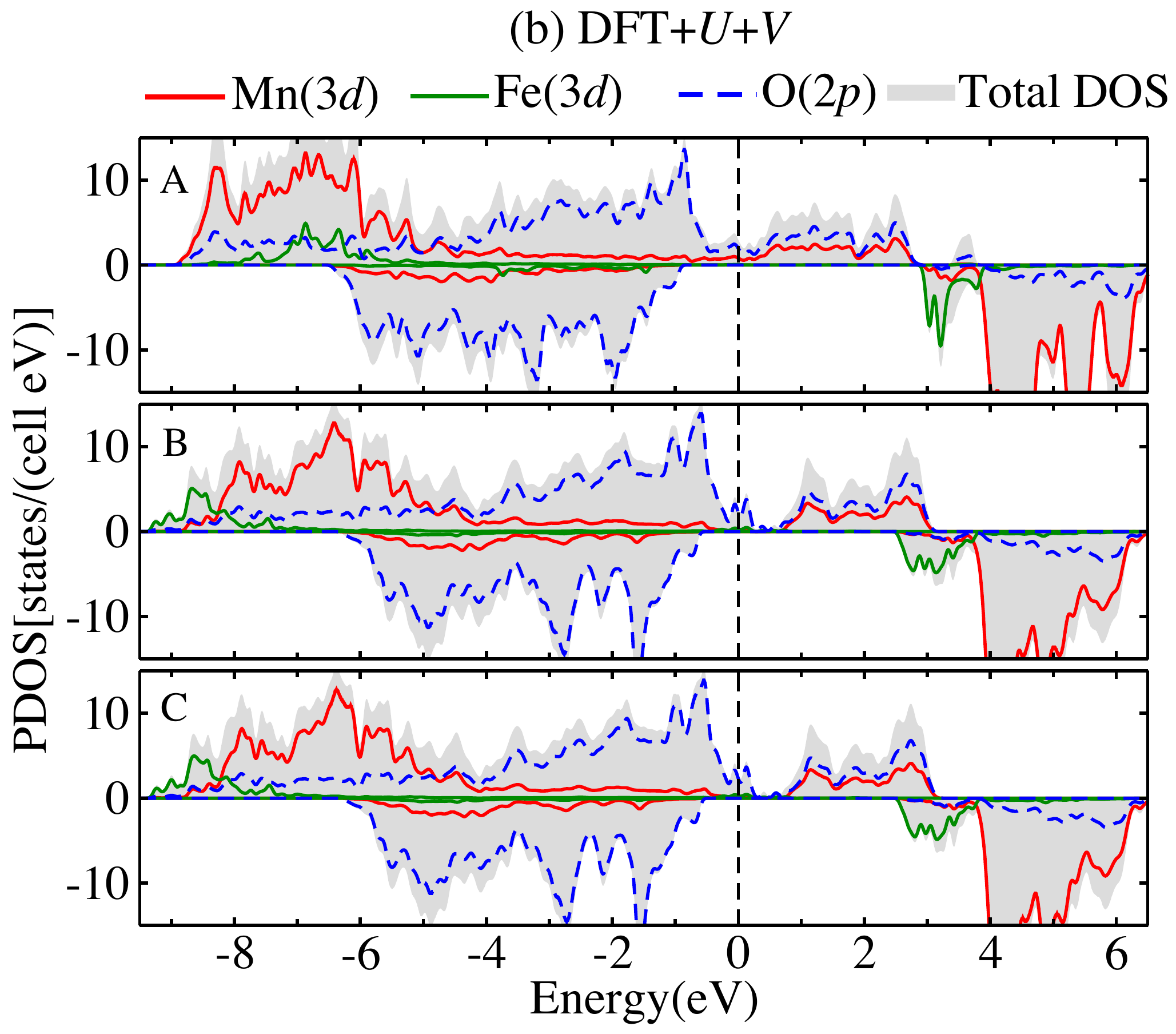}
   \caption{Spin-resolved PDOS and total DOS for three collinear magnetic orderings of the Fe-doped $\alpha$-MnO$_2$ (A, B, and C types) computed using the PBEsol functional within (a)~DFT+$U$, and (b)~DFT+$U$+$V$. For each case the Hubbard parameters $U$ and $V$ were computed using DFPT and are listed in Table~3 in the main text. Upper panels correspond to the spin-up components, while lower panels correspond to the spin-down components. The zero of energy corresponds to the Fermi energy.}
\label{fig:Fepdos_DFT}
\end{center}
\end{figure}

\begin{figure}[h!]
\begin{center}
   \includegraphics[width=0.47\linewidth]{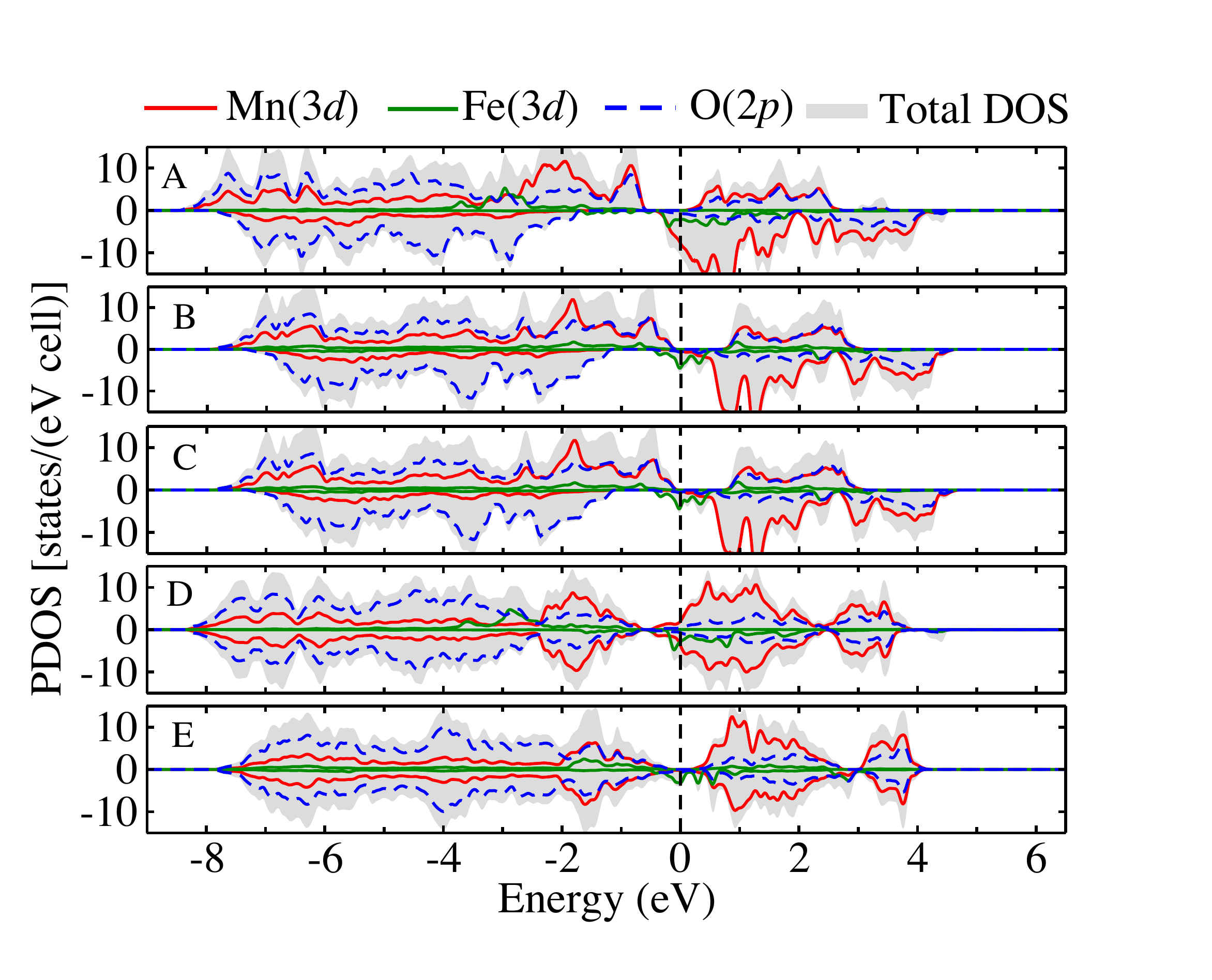} \hspace{0.4cm}
   \includegraphics[width=0.47\linewidth]{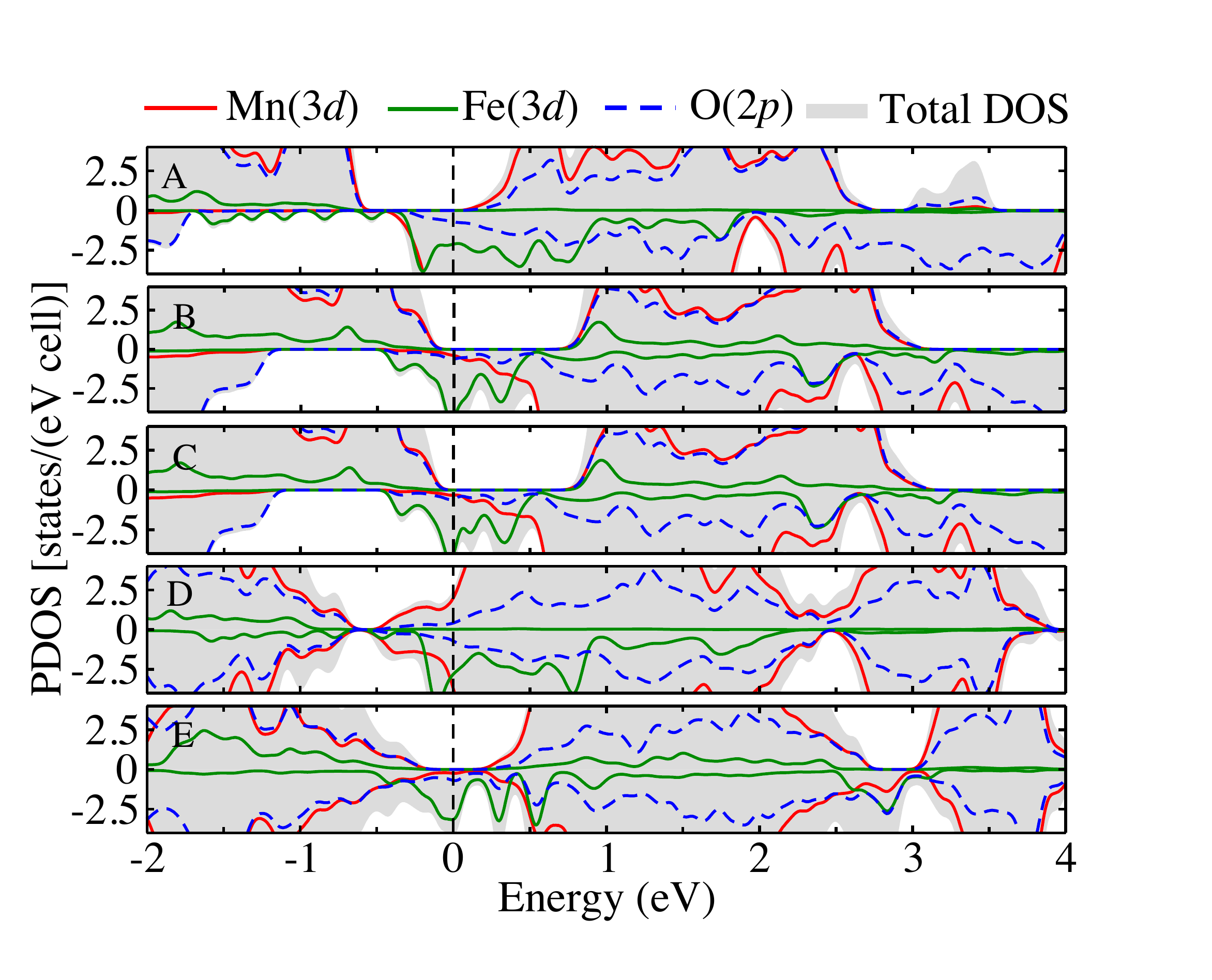}
   \caption{Spin-resolved PDOS and total DOS (left panels) and its zoom (right panels) for five collinear magnetic orderings of the Fe-doped $\alpha$-MnO$_2$ (A, B, C, D, and E) computed using standard DFT (PBEsol functional). The zero of energy corresponds to the Fermi energy.}
\label{fig:Fepdos_DFT}
\end{center}
\end{figure}

\begin{figure*}[h!]
\begin{center}
  \includegraphics[width=0.47\linewidth]{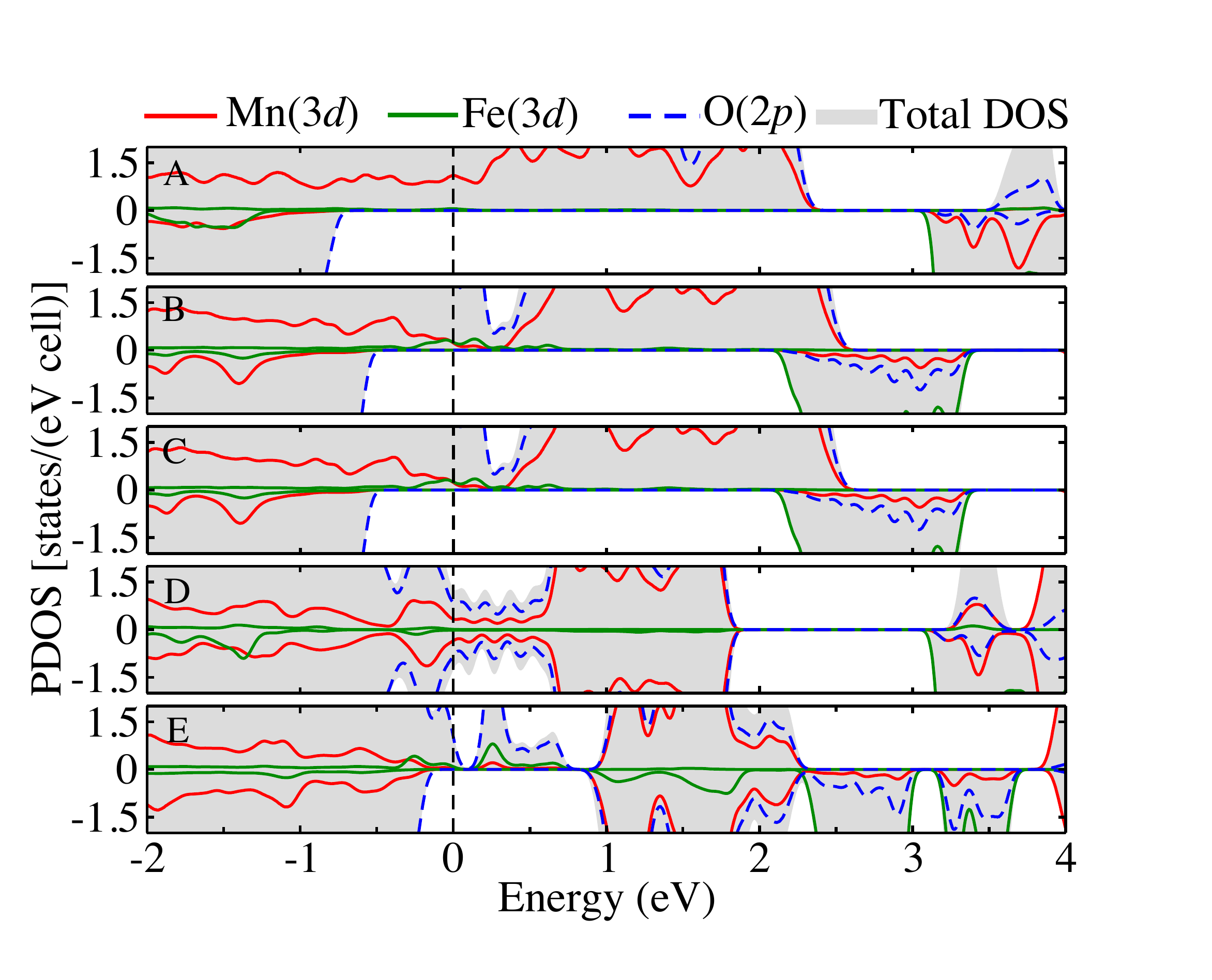}
  \hspace{0.3cm}
  \includegraphics[width=0.47\linewidth]{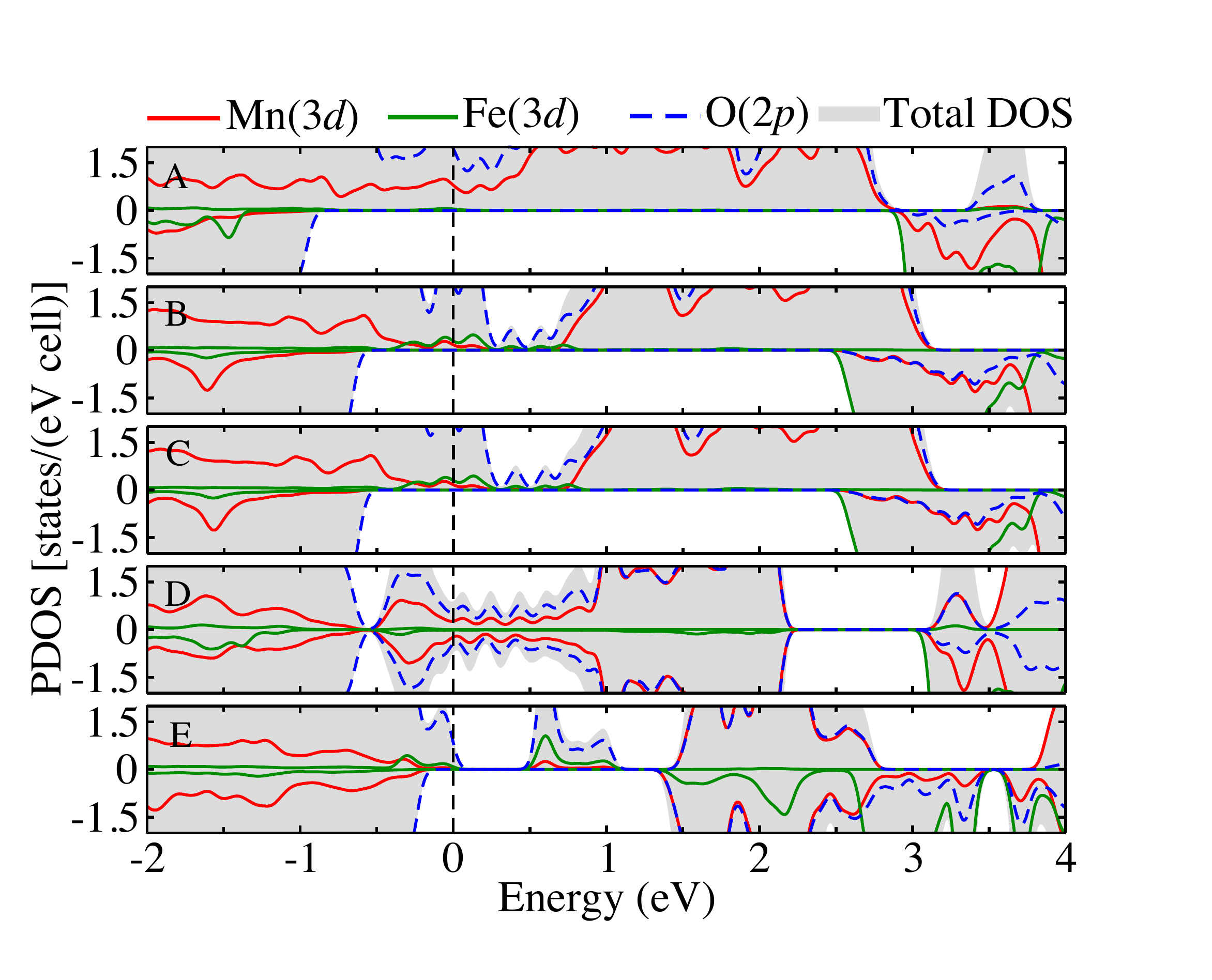}
  \caption{Zoomed-in spin-resolved PDOS and total DOS for five collinear magnetic orderings of the Fe-doped $\alpha$-MnO$_2$ (A, B, C, D, and E) computed using DFT+$U$ (left panels) and DFT+$U$+$V$ (right panels) with the PBEsol functional. For each case, the Hubbard parameters $U$ and $V$ were computed using DFPT. The zero of energy corresponds to the Fermi energy (in the case of metallic ground states) or top of the valence bands (in the case of insulating ground states).}
\label{fig:zoomedFepdos_DFT_Hubbard}
\end{center}
\end{figure*}

\begin{table*}[t]
\centering
\resizebox{\columnwidth}{!}{\begin{tabular}{c|c|ccccc|ccccc|c|c}
\hline\hline
Doping & Element & {$\lambda_1^{\uparrow}$} & {$\lambda_2^{\uparrow}$} & 
{$\lambda_3^{\uparrow}$} & 
{$\lambda_4^{\uparrow}$} & 
{$\lambda_5^{\uparrow}$} & 
{$\lambda_1^{\downarrow}$} & 
{$\lambda_2^{\downarrow}$} & 
{$\lambda_3^{\downarrow}$} & 
{$\lambda_4^{\downarrow}$} & 
{$\lambda_5^{\downarrow}$} & {$n$} &  {OS} \\
\hline
\parbox[t]{7mm}{\multirow{9}{*}{\rotatebox[origin=c]{90}{D type}}} 
  & Mn1 &      0.57 &       0.61  & {\bf 0.99} & {\bf 1.00} & {\bf 1.00} & 0.06 & 0.09 &      0.11  &      0.27  &      0.29  & 4.99 & +4 \\
  & Mn2 &      0.55 &       0.86  & {\bf 1.00} & {\bf 1.00} & {\bf 1.00} & 0.04 & 0.07 &      0.09  &      0.18  &      0.25  & 5.01 & +4 \\
  & Mn3 &      0.55 &       0.86  & {\bf 1.00} & {\bf 1.00} & {\bf 1.00} & 0.04 & 0.07 &      0.09  &      0.18  &      0.25  & 5.01 & +4 \\
  & Mn4 &      0.57 &       0.61  & {\bf 0.99} & {\bf 1.00} & {\bf 1.00} & 0.06 & 0.09 &      0.11  &      0.27  &      0.29  & 4.99 & +4 \\
  & Mn5 &      0.06 &       0.10  &      0.11  &      0.28  &      0.30  & 0.56 & 0.61 & {\bf 0.99} & {\bf 1.00} & {\bf 1.00} & 5.00 & +4 \\
  & Mn6 &      0.04 &       0.07  &      0.09  &      0.18  &      0.25  & 0.55 & 0.85 & {\bf 1.00} & {\bf 1.00} & {\bf 1.00} & 5.01 & +4 \\
  & Mn7 &      0.06 &       0.10  &      0.11  &      0.28  &      0.30  & 0.56 & 0.61 & {\bf 0.99} & {\bf 1.00} & {\bf 1.00} & 5.00 & +4 \\
  & Mn8 &      0.04 &       0.07  &      0.09  &      0.18  &      0.25  & 0.55 & 0.85 & {\bf 1.00} & {\bf 1.00} & {\bf 1.00} & 5.01 & +4 \\
  & Fe  & {\bf 0.97} & {\bf 0.99} & {\bf 1.00} & {\bf 1.00} & {\bf 1.00} & 0.02 & 0.02 &      0.03  &      0.15  & {\bf 0.94} & 6.12 & +2 \\
  \hline
\parbox[t]{7mm}{\multirow{9}{*}{\rotatebox[origin=c]{90}{E type}}} 
  & Mn1 &      0.58 &       0.59  & {\bf 0.99} & {\bf 1.00} & {\bf 1.00} & 0.07 & 0.10 &      0.11  &      0.28  &      0.29  & 5.00 & +4 \\
  & Mn2 &      0.58 &       0.59  & {\bf 0.99} & {\bf 0.99} & {\bf 1.00} & 0.07 & 0.10 &      0.11  &      0.28  &      0.29  & 5.00 & +4 \\
  & Mn3 &      0.58 &       0.59  & {\bf 0.99} & {\bf 1.00} & {\bf 1.00} & 0.07 & 0.10 &      0.11  &      0.28  &      0.29  & 5.00 & +4 \\
  & Mn4 &      0.07 &       0.09  &      0.11  &      0.28  &      0.29  & 0.58 & 0.60 & {\bf 0.99} & {\bf 1.00} & {\bf 1.00} & 5.00 & +4 \\
  & Mn5 &      0.07 &       0.10  &      0.11  &      0.29  &      0.29  & 0.58 & 0.59 & {\bf 0.99} & {\bf 1.00} & {\bf 1.00} & 5.00 & +4 \\
  & Mn6 &      0.07 &       0.10  &      0.11  &      0.28  &      0.29  & 0.58 & 0.59 & {\bf 0.99} & {\bf 1.00} & {\bf 1.00} & 5.00 & +4 \\
  & Mn7 &      0.07 &       0.10  &      0.11  &      0.28  &      0.29  & 0.58 & 0.59 & {\bf 0.99} & {\bf 1.00} & {\bf 1.00} & 5.00 & +4 \\
  & Fe  &      0.78 &  {\bf 0.99} & {\bf 1.00} & {\bf 1.00} & {\bf 1.00} & 0.09 & 0.12 &      0.18  &      0.30  &      0.37  & 5.80 & +4 \\
\hline
\end{tabular}}
\caption{Population analysis data for the $3d$ shell of Mn and Fe atoms in the Fe-doped $\alpha$-MnO$_2$ computed using DFT+$U$+$V$ (PBEsol functional). Two doping types are considered: interstitial (D) and substitutional (E). This table shows the eigenvalues of the site-diagonal occupation matrix for the spin-up ($\lambda_i^\uparrow$, $i=\overline{1,5}$) and spin-down ($\lambda_i^\downarrow$, $i=\overline{1,5}$) channels, L\"owdin occupations $n = \sum_i (\lambda_i^\uparrow + \lambda_i^\downarrow)$, and the oxidation state (OS). For the sake of simplicity we dropped the atomic site index $I$ from all quantities reported here. The eigenvalues are written in the ascending order (from left to right) for each spin channel. The eigenvalues written in bold are considered as being such that correspond to fully occupied states and thus are taken into account when determining the OS according to ref~\citen{Sit:2011SM}.}
\label{tab:OS_MnFe}
\end{table*}

\clearpage
\newpage


\providecommand{\latin}[1]{#1}
\makeatletter
\providecommand{\doi}
  {\begingroup\let\do\@makeother\dospecials
  \catcode`\{=1 \catcode`\}=2 \doi@aux}
\providecommand{\doi@aux}[1]{\endgroup\texttt{#1}}
\makeatother
\providecommand*\mcitethebibliography{\thebibliography}
\csname @ifundefined\endcsname{endmcitethebibliography}
  {\let\endmcitethebibliography\endthebibliography}{}

\end{document}